\documentclass[11pt]{article}
\usepackage[latin1]{inputenc}
\usepackage{t1enc}
\usepackage{dcolumn}
\usepackage{lscape}
\usepackage{setspace}
\usepackage{lscape}
\usepackage{amsmath,longtable,multicol,dcolumn,tabularx,graphicx,amssymb}
\usepackage{exscale,amsthm}
\usepackage{dsfont}
\usepackage{xcolor}
\usepackage{comment}

\begin{document}
\renewcommand{\baselinestretch}{1.0}
\newcolumntype{.}{D{.}{.}{-1}}
\title{The Gibbs Sampler with Particle Efficient Importance  Sampling for State-Space Models\footnote{An earlier version of the paper circulated
under the title ``Bayesian Analysis in Non-linear Non-Gaussian State-Space Models using Particle Gibbs''.}} \vspace{22mm}
\author{ Oliver Grothe\\ {\em Institute of Operations Research, Karlsruhe Institute of Technology,
 Germany}\\[0.2cm]
 Tore Selland Kleppe\\ {\em Department of Mathematics and Physics, University of
Stavanger, Norway}\\[0.2cm]
 Roman Liesenfeld\thanks{Corresponding address: Institut f\"ur \"Okonometrie  und Statistik, Universit\"at K\"oln,
Universit\"atsstr. 22a, D-50937 Köln, Germany.  Tel.: +49(0)221-470-2813;
fax: +49(0)221-470-5074. {\sl E-mail address:}
liesenfeld@statistik.uni-koeln.de (R.~Liesenfeld)}\\ {\em  Institute  of Econometrics  and Statistics, University of Cologne, Germany}\\[0.0cm]
}
\date{(\today)}
\maketitle \vspace*{-0.5cm}
\renewcommand{\baselinestretch}{1.5}
\large
\normalsize

\begin{abstract}
\begin{small}
We consider Particle Gibbs (PG) as a tool for Bayesian analysis of non-linear non-Gaussian state-space models. PG is a Monte Carlo (MC) approximation of the standard Gibbs procedure which uses   sequential MC (SMC) importance sampling inside the Gibbs procedure to update the latent and potentially high-dimensional state trajectories.
We propose to combine PG with a generic and easily implementable SMC approach known as Particle Efficient Importance Sampling (PEIS). By using  SMC importance sampling densities which are  approximately fully globally adapted to the targeted density of the states, PEIS can substantially improve the mixing and the  efficiency of the PG draws from the posterior of the states and the parameters relative to existing PG implementations.
The efficiency gains achieved by PEIS  are illustrated in PG applications to a univariate stochastic volatility model for asset returns, a non-Gaussian nonlinear local-level model for interest rates, and  a multivariate stochastic volatility model for the realized covariance matrix of asset returns.
\end{small}
\end{abstract}
\begin{small}
\vspace*{0.5cm} {\sl JEL classification: C11; C13; C15; C22.} 
\end{small}
\\

\thispagestyle{empty}
\newpage
\setcounter{section}{0}
\setcounter{equation}{0}
\setcounter{page}{1}
\renewcommand{\baselinestretch}{1.4}
\large
\normalsize

\textwidth 15.6cm \textheight 24.8cm \topmargin -0.9in

\doublespacing

\section*{1.~Introduction}
State space models (SSM) are a popular class of dynamic models used to analyze time series. In the context of SSMs, a latent Markov state variable $x_t$ $(t=1,\ldots, T)$  is observed through a response variable $y_t$, where it is  assumed that the $y_t$'s are conditionally independent given the $x_t$'s. The  measurement density for $y_t$ and  the transition density for $x_t$, depending on a vector of parameters $\theta$  are written as
\begin{equation}\label{eq:ssm}
y_t|x_t\sim g_\theta(y_t|x_t) \qquad \mbox{and} \qquad  x_t|x_{t-1}\sim f_\theta(x_t|x_{t-1}),\quad x_1\sim f_\theta(x_1),
\end{equation}
respectively. Bayesian inference about  $\theta$ and  $x_{1:T}$ relies on their joint posterior $p(\theta,x_{1:T}|y_{1:T})$, where we have used the notation $z_{s:t}$ to denote $(z_s,z_{s+1},\ldots,z_t)$. The marginal posterior for the parameters is $p(\theta|y_{1:T})\propto p_{\theta}(y_{1:T})p(\theta)$, where $p(\theta)$ denotes the prior assigned to $\theta$ and  $p_{\theta}(y_{1:T})$  the marginal likelihood.
For non-linear, non-Gaussian models, both the joint posterior of $\theta$ and $x_{1:T}$ as well as the marginal posterior for $\theta$
are analytically intractable so that inference requires approximation techniques.

A new and easy to implement tool for approximating the joint posterior $p(\theta,x_{1:T}|y_{1:T})$ in SSMs are the Particle Markov Chain Monte Carlo (PMCMC) algorithms  developed by Holenstein (2009) and Andrieu et al.~(2010), which  combine  Markov Chain Monte Carlo (MCMC) procedures with sequential Monte Carlo (SMC) algorithms. The latter  are simulation devices for  approximating  target densities such as the conditional posterior of the states  in SSMs denoted by $p_\theta(x_{1:T}|y_{1:T})$. More specifically, SMCs simulate  $x_{1:t}$-values (particles), that evolve towards the target distribution according to a combination of sequentially importance sampling (IS) and resampling.  Standard SMC implementations rely upon locally designed IS densities approximating sequentially  period by period the corresponding subcomponents of the full target density (Gordon et al., 1993, Pitt and Shephard, 1999, and Doucet and Johansen, 2009). Within the PMCMC approach such SMCs are used in order to design proposal densities for Metropolis-Hastings (MH) updates  producing MCMC draws from the respective target density.

A PMCMC algorithm available for a full Bayesian analysis in SSMs is the Particle Gibbs (PG) (Andrieu et al., 2010). It represents an MC approximation of an `ideal' (but infeasible) Gibbs algorithm that updates the joint posterior $p(\theta,x_{1:T}|y_{1:T})$ by alternately sampling from the
conditional posterior of the states $p_\theta(x_{1:T}|y_{1:T})$ and
the  conditional posterior of the parameters denoted by $p(\theta|x_{1:T},y_{1:t})$. Within PG the output of an SMC  targeting $p_\theta(x_{1:T}|y_{1:T})$ is used to obtain MCMC updates for $x_{1:T}$. While easy to implement, the baseline PG is known to suffer in applications with large $T$ from a poor mixing
since the SMC resampling  leads to a path degeneracy of the particle system hampering a sufficiently fast exploration of the domain of $x_{1:T}$ under $p_\theta(x_{1:T}|y_{1:T})$  (Whiteley et al., 2010 and Lindsten and Schön, 2012). Existing attempts to address this  poor mixing problem are to either add  a Backward Simulation step (PGBS) (Whiteley, 2010, Whiteley et al., 2010, and Lindsten and Schön, 2012) or an Ancestor Sampling step (PGAS) (Lindsten et al., 2014) to the SMC algorithm, or  to introduce  an additional MH move to update $x_{1:T}$ (PGMH) (Holenstein, 2009, p.~35).
However, as discussed further in Section 4, the efficacy of those extensions to improve the mixing of the baseline PG critically depends on how close the underlying SMC algorithm  approximates the target $p_\theta(x_{1:T}|y_{1:T})$.

In this paper we propose  to combine PG and its PGAS and PGMH extensions with a conditional version of the  Particle Efficient Importance Sampling (PEIS) developed by Scharth and Kohn (2016).
PEIS is a `forward-looking' SMC procedure which relies on the Efficient IS (EIS) technique of Richard and Zhang (2007) for designing efficient SMC IS densities. This approach exploits that  EIS  produces
close global density approximations to  (potentially high-dimensional) target densities. Scharth and Kohn (2016) uses PEIS to obtain unbiased and accurate SMC estimates of the marginal likelihood $p_\theta(y_{1:T})$  for an MH procedure targeting directly the marginal posterior of the parameters $p_\theta(\theta|y_{1:T})$.
Here we extend the application of PEIS  and show  how it can be  adapted to be used within PG  to construct  valid high-dimensional
MCMC kernels to simulate from the joint posterior $p(\theta,x_{1:T}|y_{1:T})$.

Moreover, we complement Scharth and Kohn's (2016) discussion of the PEIS by providing further important insight into the design of the PEIS. In particular, we show that it defines an SMC where in each period $t$ both sequential IS sampling as well as resampling  are aiming at being fully globally adapted to the  entire targeted posterior $p_\theta(x_{1:T}|y_{1:T})$. Based on this insight we can show that PEIS is capable to substantially improve the mixing of the PGAS and PGMH relative to their implementations using standard locally designed SMC procedures.
Empirical  illustrations of how the PEIS improves this mixing are provided in Sections 5 and 6, where we apply   PG  to a Bayesian analysis of a univariate stochastic  volatility model (SV) for asset returns,  a univariate time-discretized square-root diffusion for interest rates, and a multivariate SV model for the  realized covariance matrices of  a vector of asset returns.

An alternative PMCMC procedure to the PG  is the particle marginal MH (PMMH) (Andrieu et al., 2010). It is an MC approximation of the `ideal' MH procedure targeting directly the marginal posterior   $p(\theta|y_{1:T})$ and marginalizes the states $x_{1:T}$  by using SMC to obtain  an unbiased MC estimate of the marginal likelihood $p_\theta (y_{1:T})$. Applications   are found in Flury and Shephard (2011) and Pitt et al.~(2012), where PMMH is implemented with standard locally designed  SMCs, and   in Scharth and Kohn (2016) who combine PMMH with PEIS. However, a potential drawback of  PMMH  is, that the design of a proposal density for the MH updates of $\theta$  can be tedious, requiring a fair amount of fine tuning, especially, when the number of parameters in $\theta$ is large. Moreover, in multivariate SSMs with high-dimensional state vectors $x_t$ a sufficiently  accurate SMC estimation of the marginal likelihood $p_\theta(x_{1:T})$  as required for PMMH can be  computationally very difficult (Flury and Shephard, 2011). The advantage of PG procedures, including our proposed PG-PEIS approach, is that they offer in many applications to bypass those problems. In particular, PG can take advantage of the fact that  sampling from  $p(\theta|x_{1:T},y_{1:t})$ is often easily feasible so that the tedious design of a proposal for the marginal MH update of  $\theta$  can be avoided. Also PG does not need to MH update $x_{1:T}$ in one block so that SMC sampling from $p_\theta(x_{1:T}|y_{1:t})$ can be partitioned into a sequence  of smaller sampling problems. This can be a partitioning  into blocks along the time dimension and/or into state components for high-dimensional state vectors $x_t$. Those potential advantages of PG in SSMs with many state components  and/or parameters are empirically illustrated in our application of the PG-PEIS to the multivariate SV model.

The paper is organized as follows: In Section 2 we outline the SMC approach and in Section 3 the baseline PG. Section 4 presents the PEIS and the PG-PEIS approach (Section 4.1) and discusses potential efficiency improvements obtained by embedding the PEIS within the PGAS (Section 4.2) and the PGMH (Section 4.3). This is illustrated in Section 5 with  a Bayesian PG analysis of a univariate  SV model and a square-root diffusion model and in Section 6 with a multivariate SV specification. Section 7 concludes.
\section*{2.~Sequential Monte Carlo (SMC)}
\subsection*{2.1~Definition of SMC}
Let $\pi(x_{1:T})$ denote  the target density to be approximated/simulated with the following sequence of intermediate target densities:
\begin{equation}\label{eq:pi_t}
\pi_t(x_{1:t})=\frac{\gamma_t(x_{1:t})}{z_t},\qquad z_{t}=\int \gamma_t(x_{1:t}) dx_{1:t},\qquad t=1,\ldots,T,
\end{equation}
with $\pi_T(x_{1:T})\equiv \pi(x_{1:T})$. In an SSM of the form given by (\ref{eq:ssm}) the full target is $\pi(x_{1:T})=p_\theta(x_{1:T}|y_{1:T})$ and for standard SMCs the intermediate targets are defined as
$\pi_t(x_{1:t})\equiv p_\theta(x_{1:t}|y_{1:t})$, so that
\begin{eqnarray}\label{eq:gamma_t}
\gamma_t(x_{1:t})&=&
p_\theta(x_{1:t},y_{1:t})= \left[\prod_{\tau=2}^t  g_\theta(y_\tau|x_\tau)f_\theta(x_\tau|x_{\tau-1})\right]g_\theta(y_1|x_1)f_\theta(x_1),\\
z_t&=& p_\theta(y_{1:t})=\int p_\theta(x_{1:t},y_{1:t})dx_{1:t},
\end{eqnarray}
where the sequence $z_t=p_\theta(y_{1:t})$ represent the marginal likelihoods.

SMC algorithms as discussed in Capp\'{e} et al.~(2007) and Doucet and Johansen (2009), consist of recursively producing, for each period $t$, a
weighted particle system $\{x_{1:t}^i, w_t^i\}_{i=1}^N$ with $N$ particles $x_{1:t}^i$  and
(non-normalized) IS weights $w_t^i$ such that 
$\pi_t(x_{1:t})$  in  (\ref{eq:pi_t}) can be  approximated by the point mass distribution
\begin{equation}\label{eq:SMC-app-pi}
\hat \pi_t(dx_{1:t})=\sum_{i=1}^N W_t^i \,\delta_{x_{1:t}^i}(dx_{1:t}), \qquad W_t^i=\frac{w_t^i}{\sum_{l=1}^N w_t^l},
\end{equation}
where $\delta_{x}(\cdot)$ denotes the Dirac delta mass located at $x$.
In period $t$, the
particle system $\{x_{1:t}^i, w_t^i\}_{i=1}^N$ is obtained from the period-$(t-1)$ system $\{x_{1:t-1}^i, w_{t-1}^i\}_{i=1}^N$ by  drawing from an IS-density $q_t(x_t|x_{1:t-1}^i)$ to propagate the inherited particles $x_{1:t-1}^i$ to $x_{1:t}^i=(x_{t}^i,x_{1:t-1}^i)$ and  updating the  IS weights according to
\begin{equation}\label{eq:SMC-is-weights}
w_t^i =W_{t-1}^i\frac{\gamma_t(x_{1:t}^i)}{\gamma_{t-1}(x_{1:t-1}^i)q_t(x_t^i|x_{1:t-1}^i)}.
\end{equation}

In most applications, the variance of the IS weights $w_t^i$
increases exponentially with $t$  reducing the  effective sample size of the particle system (an effect known as  `weight degeneracy'). Hence, SMC algorithms include a resampling step before propagating the particles  $x_{1:t-1}^i$. It consists in sampling $N$ `ancestors particles' from  $\{x_{1:t-1}^i\}_{i=1}^N$ according to their normalized  IS weights $\{ W_{t-1}^i\}$
and   setting in (\ref{eq:SMC-is-weights})  the  IS weights $W_{t-1}^i$ for the  redrawn
$x_{1:t-1}^i$-particles  all equal to $1/N$. This resampling step amounts to sampling for $t=2,...,T$ the (auxiliary) indices of the  ancestor particles $x_{1:t-1}^i$  denoted by $a_t^i$.  For a discussion of  resampling schemes including multinomial  resampling, see Doucet and Johansen (2009).

This procedure provides an approximation of the full target density $\pi(x_{1:T})$  given by $\hat \pi_T(dx_{1:T})$ according to (\ref{eq:SMC-app-pi}). Approximate samples from $\pi(x_{1:T})$  can be obtained by sampling $x_{1:T}^i\sim \hat \pi_T(dx_{1:T})$, which is done by choosing   particles $x_{1:T}^i$  according to their  probabilities $W_T^i$.
If required, the normalizing constant $z_T$ of $\pi(x_{1:T})$ is estimated by
\begin{equation}\label{eq:smc-app-integrating const}
 \hat z_T = \prod_{t=1}^T \left(\sum_{i=1}^N w_t^i\right).
\end{equation}

In the SSM context, such an SMC produces an approximation of $\pi(x_{1:T})=p_\theta(x_{1:T}|y_{1:T})$, denoted by $\hat p_\theta(x_{1:T}|y_{1:T})$,  corresponding  approximate samples $x_{1:T}^i\sim\hat p_\theta(x_{1:T}|y_{1:T})$, and an MC approximation to the full marginal likelihood $z_T= p_\theta(y_{1:T})$ written as $\hat p_\theta(y_{1:T})$. These are the main inputs of PG algorithms implemented for  Bayesian analyzes of SSMs.

\subsection*{2.2~SMC implementations in state  space models}

A critical issue in implementing an SMC is the choice of the IS densities  $q_t(x_t|x_{1:t}^i)$. The main  recommendation is to design them {\sl locally}  so as to
minimize the conditional variance of the IS weights in   (\ref{eq:SMC-is-weights}) given $x_{1:t-1}^i$. 
For  SSM applications with $\pi_t(x_{1:t})\propto p_\theta(x_{1:t},y_{1:t})$ as given by  (\ref{eq:gamma_t}), those IS weights become
\begin{equation}\label{eq:SSM-is-weights}
w_t^i =W_{t-1}^i \frac{g_\theta(y_t|x_t^i)f_\theta(x_t^i|x_{t-1}^i)}{q_t(x_t^i|x_{1:t-1}^i)}.
\end{equation}
The  most popular (but suboptimal) selection for the IS densities  are the transition densities  $f_\theta(x_t|x_{t-1}^i)$ used by the Bootstrap Particle Filter (BPF) (Gordon et al., 1993). In scenarios where the measurement density $g_\theta$ is fairly flat in $x_t$, this selection typically leads to a satisfactory performance. A selection  which sets the  variance of the IS weights in  (\ref{eq:SSM-is-weights}) {\sl conditional on}  $x_{t-1}^i$ to zero is $p_\theta(x_t|y_t,x_{t-1}^i)\propto g_\theta(y_t|x_t)f_\theta(x_t|x_{t-1}^i)$, leading to the conditionally optimal Particle Filter (Doucet and Johansen, 2009).  Further improvements can be achieved  by replacing the standard resampling schemes based on the IS weights in (\ref{eq:SSM-is-weights}) by more sophisticated ones which favor ancestor particles which will be in regions with high probability mass after their propagation. This is implemented by the Auxiliary Particle Filter (APF) (Pitt and Shephard, 1999).

In contrast to those locally designed SMCs, the `forward-looking'  PEIS  of Scharth and Kohn (2016)  uses information from future observations  and constructs  IS densities and resampling weights from a global functional approximation to the full target $\pi(x_{1:T})$. It is well understood that using future information typically improves the performance of SMC's (Lin et al., 2013). The forward-looking approach  most closely related to  PEIS is the iterated APF (iAPF) of  Guarniero et al.~(2017). Its design also  aims at constructing IS densities  from a global approximation of the full target but the way  to find this approximation is  different to that of PEIS. The iAPF embeds this approximation into  repeated SMC runs which are complemented by backward sweeps estimating the globally adapted IS densities from the particle system of the previous SMC run. In contrast, PEIS produces those estimates before running the SMC.

Irrespectively  of the particular IS density selected for an SMC implementation,  the resampling steps used to mitigate the weight degeneracy,  lead to a loss of diversity among the particles as the resultant sample may contain many repeated points. Hence, in many applications   resampling  is performed dynamically, i.e., only when the weight degeneracy exceeds a certain threshold (Doucet and Johansen, 2009).

\section*{3.~Particle Gibbs (PG)}

\subsection*{3.1~Baseline Particle Gibbs algorithm}

For a Bayesian analysis in a non-linear, non-Gaussian SSM  the `ideal' Gibbs sampler targeting the joint posterior $p(\theta, x_{1:T}|y_{1:T})$ and alternately sampling from the conditional posteriors
$p_\theta(x_{1:T}|y_{1:T})$ and $p(\theta|x_{1:T},y_{1:T})$ is typically unfeasible since exact sampling from $p_\theta(x_{1:T}|y_{1:T})$ is impossible.
The PG approach of Andrieu et al.~(2010) uses an SMC  targeting  $\pi(x_{1:T})=p_\theta(x_{1:T}|y_{1:T})$ in order to propose approximate samples from this distribution in such a way that the ideal Gibbs sampler is `exactly approximated'. This is achieved by augmenting the target density of the ideal Gibbs sampler $p(\theta,x_{1:T}|y_{1:T})$ to include {\sl all} the  random variables which are produced by the SMC
in order to generate an $x_{1:T}$-proposal $x_{1:T}^k$.
The set of those SMC random variables is given by ($\bar x_{1:T}$, $\bar a_{1:T-1}$,$k$), where $\bar x_{1:T}=(x_{1:T}^1,\ldots,x_{1:T}^N)$ denotes the $N$ particle trajectories,
$\bar a_{1:T-1}=(\bar a_1,\ldots,\bar a_{T-1})$ with  $\bar a_{t}=(a_t^1,\ldots, a_t^N)$ the $N(T-1)$ ancestor indices generated by the resampling steps, and $k$ the particle  index  drawn according to the SMC weights $\{W_T^i\}$. In order to keep track of the ancestor indices of particle $x_{1:T}^k$, we use the variable  $b_t^k$ 
to denote the index of the ancestor particle of $x_{1:T}^k$ at generation $t$. It obtains recursively as  $b_T^k=k$ and $b_t^k=a_t^{b_{t+1}^k}$, and with the resulting sequence $b_{1:T}^k=(b_1^k,\ldots,b_T^k)$  we  can write $x_{1:T}^k=(x_1^{b_1^k},\ldots,x_T^{b_T^k})$.
The PG then obtains as a standard Gibbs sampler for the augmented target density over $\theta,k,\bar x_{1:T}, \bar a_{1:T-1}$.

The Gibbs sampler for this augmented target density requires a special type of SMC, referred to as {\sl conditional} SMC, where one of the particles  $\{x_{1:T}^i\}_{i=1}^N$ is specified a-priori. This pre-specified reference particle denoted by $x_{1:T}'$ is then retained throughout the entire SMC sampling process. To accomplish this  for a multinomial resampling scheme, one can set $x_t^{1}\equiv x_{t}'$ and $a_t^1\equiv 1$  for all periods  and use the SMC to sample the $x_t^i$'s and $a_t^i$'s only for $i=2,..., N$  (Lindsten et al., 2014). This produces a set of $N$ particles and IS weights $\{x_{1:T}^i, w_T^i \}_{i=1}^N$, where the first particle coincides with the pre-specified one, i.e., $x_{1:T}^{1}=x_{1:T}'$. Based on such a conditional SMC the PG algorithm for an SSM is given by:

\onehalfspacing
\begin{itemize}
\item[]{\bf \underline {PG algorithm}}
\begin{itemize}
\item[(i)] {\it Initialization $(j=0)$: Set arbitrarily $\theta^{(0)}$, run an SMC targeting $p_{\theta^{(0)}}(x_{1:T}|y_{1:T})$,
       and sample $x_{1:T}^{(0)}\sim \hat p_{\theta^{(0)}}(x_{1:T}|y_{1:T})$   by drawing a particle index $k$ with Pr$(k=i)=W_T^i$ and setting $x_{1:T}^{(0)}=x_{1:T}^k$.}
\item[(ii)] {\it For iteration $j\geq 1$:}
\begin{itemize}
\item[-]{\it  sample  $\theta^{(j)}\sim p(\theta| x_{1:T}^{(j-1)},y_{1:T} )$ },
\item[-]{\it  run a conditional SMC targeting  $p_{\theta^{(j)}}(x_{1:T}|y_{1:T})$  conditional on $x_{1:T}^{(j-1)}$,
       and sample $x_{1:T}^{(j)}\sim \hat p_{\theta^{(j)}}(x_{1:T}|y_{1:T})$  by drawing a particle index $k$ with Pr$(k=i)=W_T^i$ and setting $x_{1:T}^{(j)}=x_{1:T}^k$.}
\end{itemize}
\end{itemize}
\end{itemize}

\doublespacing

The Markov kernel defined by this PG algorithm admits
\begin{equation}\label{extended PG target}
\tilde p(\theta,k, \bar x_{1:T}, \bar a_{1:T-1} )=\frac{1}{N^T}p(\theta,x_{1:T}^k|y_{1:T}) \tilde p( \bar x_{1:T}^{\backslash b_{1:T}^k}, \bar a_{1:T-1}^{\backslash b_{1:T}^k}| x_{1:T}^k,b_{1:T}^k,\theta)
\end{equation}
as invariant density, where $\tilde p$ on the r.h.s.~represents the density of the random variables generated by the conditional SMC given $x_{1:T}^k,b_{1:T}^k,\theta$. Here we have used the notation $ \bar x_{1:T}^{\backslash b_{1:T}^k}$ to denote the set of $N$ particles $\bar x_{1:T}$ excluding $x_{1:T}^k$, and $\bar a_{1:T}^{\backslash b_{1:T}^k}$ the set of all ancestor indices excluding  those associated with $x_{1:T}^k$. (The specific form of this conditional density is found in Andrieu et al., 2010, Section 4.4).
The first two factors of the PG target in  (\ref{extended PG target}) represent the marginal density of $\theta$, $x_{1:T}^k$, $b_{1:T}^k$,
\begin{equation}
\tilde p(\theta,x_{1:T}^k, b_{1:T}^k)= \frac{1}{N^T}p(\theta,x_{1:T}^k|y_{1:T}),
\end{equation}
defined to be the original target of interest $p(\theta,x_{1:T}^k|y_{1:T})$ up to the factor $1/N^T$ representing a discrete uniform density over the index variables in $b_{1:T}^k$. It follows that the Markov kernel defined by the PG leaves the original target $p(\theta,x_{1:T}^k|y_{1:T})$ invariant and delivers under weak regularity conditions  a sequence of draws $\{\theta^{(j)},x_{1:T}^{(j)} \}$ whose marginal distribution
converge for any $N>1$ to $p(\theta,x_{1:T}|y_{1:T})$ as $j\to\infty$ (Andrieu et al., 2010, Theorem 5).

Existing PG applications  use locally designed SMCs like the BPF  with  resampling steps  performed at every  period $t$. Dynamic  resampling, while  possible,
is computationally inefficient since the conditional SMC at  PG-iteration step $j$ requires simulating a set of $N-1$ particles not only consistent with the retained path $x_{1:T}'=x_{1:T}^{(j-1)}$  but  also with the resampling times of the SMC pass which has produced the retained path (Holenstein, 2009, Section, 3.4.1).

\subsection*{3.2~Particle Gibbs and the SMC-path degeneracy}
The baseline PG will, if implemented using SMCs with resampling steps at every period $t$, have a very  poor mixing,
especially, when $T$ is large (Whiteley et al., 2010). The reason for this is that  the SMC resampling
inevitably leads to a path degeneracy  of the particle system. This means that every period-$t$  resampling step  will sequentially reduce  for a fixed $s<t$ and increasing $t$ the number of unique particle  values representing $x_{1:s}$, which progressively  reduces the quality  of the SMC samples for the path $x_{1:t}$  under $p_\theta(x_{1:t}|y_{1:t})$. The consequence of this SMC path degeneracy for the PG is that  at iteration step $j$ the new trajectory $x_{1:T}^{(j)}$ tend to coalesce (when moving from period $T$ backwards to period 1) with the previous one $x_{1:T}^{(j-1)}$  which is retained as the reference particle $x_{1:T}'$  throughout conditional SMC sampling. Thus, the resulting particle system degenerates towards this `frozen' path,  leading to a  highly dependent Markov chain.

Before we discuss in the next section solutions to this problem of the baseline PG, two  points bear mentioning. First, it is not the SMC path degeneracy {\sl per se} which leads to the poor mixing of the PG, but the degeneration  of the particle system towards the retained conditional SMC reference particle $x_{1:T}'$. On the other hand, however, SMC implementations addressing successfully the path degeneracy problem can be used to fight the poor mixing of the PG. Second, by construction {\sl any} SMC, whether implemented using locally or globally optimal IS densities,  will lead to a fast degeneration of the SMC paths, when resampling is performed  every period. This precludes that the mixing problem of the baseline  PG
can be successfully addressed  solely by the design of the SMC IS densities and resampling schemes.

\section*{4. Extensions of the baseline Particle Gibbs}
In order to address the mixing problem  of the baseline PG caused by the SMC path degeneracy the following strategies have been proposed: The first one is to augment the baseline PG by an additional particle MH update step (PGMH) proposing at each PG-iteration step $j$  a completely new SMC path for $x_{1:T}$
(Holenstein, 2009, Section 3.2.3). A second alternative is to add additional Ancestor Sampling (AS) steps to the conditional SMC (PGAS), which assign at each period $t$ a new artificial $x_{1:t-1}$-history to the partial frozen path $x_{t:T}'$ (Lindsten et al., 2014). Yet another strategy is to add to the conditional SMC a backward simulation step (PGBS) based on the output of the SMC forward filtering pass (Whiteley, 2010, Whiteley et al., 2010, and Lindsten and Schön, 2012). However,  this approach is in Markovian SSMs  probabilistically  equivalent to the PG with ancestor sampling  (Lindsten et al., 2014).

As illustrated in our applications below the efficacy of the PGAS   and  PGMH to improve the mixing of the baseline PG critically depends on the SMC  which is used for their implementation. In particular, an efficient PGMH implementation  requires for the additional MH step  numerically precise SMC estimates of the marginal likelihood $p_\theta(y_{1:T})$, which can  in high-dimensional applications be  too much of a challenge for locally designed SMCs.     On the other hand, the efficacy of the PGAS's ancestor sampling to improve the  mixing can be seriously hampered by a large inequality of the IS weights $w_t^i$, which is to be expected for local SMCs, especially, in SSM applications with measurement densities which are very informative about the states and poorly aligned with the states' predictive density.
Since, as  mentioned above and further detailed below, the PEIS of Scharth and Kohn (2016)   uses  IS densities  obtained from a global minimization of the variance of the IS weights $w_t^i$ across all periods aiming at producing very close SMC approximations to  $p_\theta(x_{1:T}|y_{1:T})$ and $p_\theta(y_{1:T})$, we propose to use this PEIS in order to improve the efficiency of the PGAS and PGMH.

The extensions of the baseline PG outlined above are detailed in the next sections: In Section 4.1 we describe the PEIS. In Sections 4.2 and 4.3 we present the PGAS and PGMH, respectively, and discuss the potential efficiency improvements obtained if they are implemented with the PEIS.

\subsection*{4.1 Particle EIS (PEIS)}
\subsubsection*{4.1.1 EIS principle}
The PEIS of Scharth and Kohn (2016) is an  SMC which
uses Richard and Zhang's (2007)  EIS procedure   to design  IS densities as well as a resampling scheme. EIS  is a generic algorithm  which aims at constructing a global IS density $q$ for $x_{1:T}$,  which  approximates  $p_\theta(x_{1:T}|y_{1:T})\propto p_\theta(x_{1:T},y_{1:T})$ as close as possible by minimizing the variance of the resulting IS ratio.
For this, the  global IS density is factorized conformably with $p_\theta(x_{1:T},y_{1:T})$ in (\ref{eq:gamma_t}) into
\begin{equation}
q(x_{1:T}; c)=\left[\prod_{t=2}^T q_t(x_t|x_{t-1};c_t)\right]q_1(x_1;c_1),
\end{equation}
with
\begin{equation}\label{eq:q_t_EIS}
q_t(x_t|x_{t-1};c_t) =\frac{k_t(x_{t},x_{t-1};c_t)}{\chi_t(x_{t-1};c_t)},\qquad \chi_t(x_{t-1};c_t)=\int k_t(x_{t},x_{t-1};c_t) dx_t,
\end{equation}
where ${\cal K}=\{ k_t(\cdot ;c_t), c_t \in \mathcal{C}_t \}$ is a preselected parametric class of density kernels indexed by a
vector of (auxiliary) parameters $c_t$  and with  point-wise computable integrating factors given by $\chi_t$. For any given $c=(c_1,...,c_T)$  the global IS ratio $p_\theta(x_{1:T},y_{1:T})/q(x_{1:T}; c)$ can be sequentially factorized so as to obtain
\begin{equation}\label{eq:global-IS-ratios}
\frac{p_\theta(x_{1:T},y_{1:T})}{q(x_{1:T}; c)}=\chi_1(c_1)\prod_{t=1}^T \left[\frac{ g_\theta(y_t|x_t)f_\theta(x_t|x_{t-1})\chi_{t+1}(x_{t};c_{t+1})}{k_t(x_{t},x_{t-1};c_t)}  \right],\qquad \chi_{T+1}(\cdot)\equiv 1.
\end{equation}
This implies that for a sequence of density kernels $k_t^*(.;c_t)$  with  parameter values $c_t^*$ and integrating factors $\chi_t^*(\cdot;c_t^*)$,  satisfying the back-recursive sequence of identities
\begin{equation}\label{eq:opt_k}
k_t^*(x_{t},x_{t-1};c_t^*)= g_\theta(y_t|x_t)f_\theta(x_t|x_{t-1})\chi^*_{t+1}(x_{t};c^*_{t+1}),\qquad t= T,\ldots,1,\quad \chi_{T+1}^*(\cdot)\equiv 1,
\end{equation}
the variance of the global IS ratio (\ref{eq:global-IS-ratios}) is zero and the resulting sequence of IS densities $q_t^*(x_t|x_{t-1};c_t^*)=k_t^*(x_{t},x_{t-1};c_t^*)/\chi^*_{t}(x_{t-1};c^*_{t})$ define a joint density  $q^*(x_{1:T};c^*)$ which is equal to the target $p_\theta(x_{1:T}|y_{1:T})$. 
Except for Gaussian linear SSMs
the construction of  those optimal IS densities is infeasible  since the sequence of integrating factors $\chi_t^*$ are analytically intractable integrals.

In order to obtain period by period close functional  approximations to the optimal IS density kernels $k_t^*$ and their integrating factors $\chi_t^*$, EIS solves for  the preselected  parametric class of kernels ${\cal K}$
the following back-recursive sequence of least squares (LS) approximation problems:
\begin{equation}\label{eq:EIS-LS}
(\hat c_t,\hat \alpha_t)=\arg\min_{c_t\in {\cal C}_t,\alpha_t\in \mathds{R}}\sum_{i=1}^R\left\{ \ln\left[ g_\theta(y_t|x_t^i)f_\theta(x_t^i|x_{t-1}^i)\chi_{t+1}(x_{t}^i;\hat c_{t+1})\right]\right. \qquad\qquad\qquad
\end{equation}
\[
\qquad\qquad\qquad\qquad\qquad\left.-\alpha_t-\ln k_t(x_{t}^i,x_{t-1}^i;c_t) \right\}^2,\qquad t=T,\ldots,1,
\]
where  $\alpha_t$ represents an intercept, and $\{x_{1:T}^i\}_{i=1}^R$ denote $R$ independent trajectories drawn from $q(x_{1:T}; c)$ itself. Thus, $\hat c$ results as a fixed point solution to the sequence $\{\hat c^{[0]},\hat c^{[1]},\ldots \}$ in which $\hat c^{[\ell]}$ is obtained from (\ref{eq:EIS-LS}) under trajectories drawn from $q(\cdot; \hat c^{[\ell-1]})$. In order to ensure convergence to a fixed-point solution it is critical that all $x_{1:T}$ draws generated for the sequence $\{ \hat c^{[\ell]}\}$  be produced by using a single set of  $T\cdot R$ canonical random numbers (CRNs) $\bar u_{1:T} = (u_{1:T}^1,\ldots,u_{1:T}^R)$ from a density denoted by $p(\bar u_{1:T})$, which is typically that of uniforms or
standard normals. Since the $\hat c_t$'s are implicit functions of $\theta$  maximal efficiency requires  complete reruns of the EIS regressions for any new value of $\theta$.

The selection of the parametric class ${\cal K}$ of kernels $k_t$ is inherently model-specific since they are meant to provide a functional approximation to the product $g_\theta f_\theta \chi_{t+1}$. General guidelines for the selection of ${\cal K}$
are found in Richard and Zhang (2007). If $k_t$ is constructed  using kernels within the  exponential family of distributions the EIS
LS regressions (\ref{eq:EIS-LS}) take the form of simple {\sl linear} LS problems. This together with the property that such exponential kernels are closed under multiplication simplifies the application of EIS to the SSMs we consider below. There we use for $k_t$ Gaussian kernels as well as kernels for mixtures of Gaussian distributions.

\subsubsection*{4.1.2 PEIS as an APF aiming at full global adaption}
The PEIS is an SMC, which is constructed from the output of the EIS algorithm  consisting of $\{k_t(\cdot; \hat c_t),$ $\chi_t(\cdot; \hat c_t)\}$ as follows: Firstly, it makes use of the APF principle (Pitt and  Shephard, 1999) and replaces
the standard resampling scheme based upon the IS weights in  (\ref{eq:SSM-is-weights}) by  a scheme, which favors particles that are more likely to survive  the next resampling steps. This can be implemented  within the standard SMC  outlined in Section 2.1,
by replacing the natural intermediate targets $\pi_t(x_{1:t})$  in (\ref{eq:gamma_t}) by auxiliary targets, which
include  information of future $y_t$-measurements (Doucet and Johansen, 2009, Section 4.2).
The particular auxiliary targets used by  PEIS  are given by
\begin{equation}\label{eq:pi_t-EIS-1}
\pi_t(x_{1:t})\propto \gamma_t(x_{1:t})\equiv p_\theta(x_{1:t},y_{1:t})\chi_{t+1}(x_{t};\hat c_{t+1}). 
\end{equation}
Secondly, PEIS uses as SMC-IS densities the ones obtained from the EIS auxiliary regressions (\ref{eq:EIS-LS})
\begin{equation}\label{eq:IS-density-EIS-1}
 q_t(x_t|x_{1:t-1})\equiv q_t(x_t|x_{t-1};\hat c_{t})=\frac{k_t( x_{t},  x_{t-1};\hat c_t)}{\chi_{t}(x_{t-1};\hat c_{t}) }.
\end{equation}
Under those PEIS selections given in  (\ref{eq:pi_t-EIS-1})  and (\ref{eq:IS-density-EIS-1}), the SMC IS weights in (\ref{eq:SMC-is-weights}) become
\begin{equation}\label{PEIS-is-weights}
w_t^i=W_{t-1}^i\frac{g_\theta(y_t|x_t^i)f_\theta(x_t^i|x_{t-1}^i)\chi_{t+1}(x_{t}^i;\hat c_{t+1})}{k_t(x_{t}^i,x_{t-1}^i;\hat c_t)}.
\end{equation}

To motivate this PEIS, we note that if the auxiliary targets in (\ref{eq:pi_t-EIS-1}) and the IS densities in  (\ref{eq:IS-density-EIS-1}) were constructed using the kernels $k_t^*(x_{t},x_{t-1};c_t^*)$ and their integrating factors $\chi_{t+1}^* (x_{t};c_{t+1}^*)$ defined by recursion (\ref{eq:opt_k}), then the corresponding SMC IS weights $w_t^i$ in  (\ref{PEIS-is-weights}) would become $w_t^i=W_{t-1}^i\cdot 1$ so that their variance would be zero each for each period $t$. Hence, this (generally infeasible) SMC based on  the functions $k_t^*(x_{t},x_{t-1};c_t^*)$ and  $\chi_{t+1}^*(x_{t};c_{t+1}^*)$ defines the {\sl globally fully adapted} SMC. Since the kernels $k_t( x_{t},x_{t-1};\hat c_t)$ and their integrating factors $\chi_{t+1}(x_{t};\hat c_{t+1})$ defining the PEIS
are  constructed so as to provide within the selected class of kernel functions ${\cal K}=\{k_t(\cdot;c_t), c_t \in \mathcal{C}_t  \}$ the best possible functional approximations to their optimal counterparts $k_t^*(x_{t},x_{t-1};c_t^*)$ and  $\chi_{t+1}^*(x_{t};c_{t+1}^*)$,  PEIS is  designed to get as close to the globally fully adapted SMC as possible. As such the PEIS aims at minimizing period by period the variance of the SMC IS weights $w_t^i$ and thus the SMC weight degeneracy.

To complement this motivation of the PEIS, we highlight its forward looking nature by linking the functions approximated by  $\chi_{t+1}(x_{t}; \hat c_{t+1})$ and  $k_t(x_{t},x_{t-1};\hat c_t)$ to the corresponding `look ahead' densities of the SSM. Those links obtain from recursion (\ref{eq:opt_k}), which  implies   that
\begin{equation}\label{EIS-chi}
\chi_{t+1}^*(x_{t} ;  c_{t+1}^*) = p_\theta(y_{t+1:T}|x_{t}) \quad\mbox{and}\quad  q_t^*(x_t|x_{t-1}; c_{t}^*) = \frac{k_t^*(x_{t},x_{t-1};c^*_t)}{\chi_{t}^*(x_{t-1}; c^*_t)}= p_\theta(x_{t}|x_{t-1}, y_{t:T}),
\end{equation}
where $p_\theta(y_{t+1:T}|x_{t})$ is the multiperiod a-head predictive likelihood and $p_\theta(x_{t}|x_{t-1}, y_{t:T})$ the associated conditional predictive posterior for $x_t$ given $x_{t-1}$ (for the  derivations of these identities, see the online appendix.) Equation (\ref{EIS-chi}) implies that  the auxiliary SMC target of the PEIS in (\ref{eq:pi_t-EIS-1}) includes  by $\chi_{t+1}(x_{t};\hat c_{t+1})$ an  approximation of
$p_\theta(y_{t+1:T}|x_t)$
so that the resulting resampling scheme based on the IS weights (\ref{PEIS-is-weights}) favors ancestor particles with high probability masses in all subsequent periods.
Simultaneously, the PEIS sampling density in (\ref{eq:IS-density-EIS-1}) proposes, as an approximation to  $p_\theta(x_{t}|x_{t-1}, y_{t:T})$, period by period $x_t$-particles  which are adapted in light of  the current and all subsequent measurements $y_{t:T}$.
Thus,  both PEIS sampling and resampling are in each period globally adapted to the final target $p_\theta(x_{1:T}|y_{1:T})$.

Since  PEIS  requires to run the sequence of $T$ auxiliary regressions (\ref{eq:EIS-LS}) {\sl before} producing via the sequence  of SMC steps a weighted particle system $\{x_{1:T}^i,w_{T}^i\}$, its design aiming at perfect global adaption comes at additional computational costs relative to   standard locally designed SMC procedures. However, as illustrated  in the applications  below, the substantial improvements of the approximation  to $p_\theta (x_{1:T}|y_{1:T})$  gained by the PEIS may  outweigh its additional computational costs.
In the following algorithm we provide a pseudo code of the PEIS.

\onehalfspacing
\begin{itemize}
\item[]{\bf \underline {PEIS  algorithm}}
\begin{itemize}
\item[(i)] {\it  Draw a set of CRNs $\bar u_{1:T}$ and compute $\hat c = (\hat c_1,\ldots,\hat c_T)$ by iteratively  drawing from  $q(x_{1:T};\hat c^{[\ell]})$ and producing $c^{[\ell+1]}$ via the $T$ auxiliary EIS regressions in  (\ref{eq:EIS-LS}), and store $\hat c$.}
\item[(ii)] {\it For $t=1$:}
\begin{itemize}
\item[-]{\it Sample $x_1^i\sim q_1(x_1,\hat c_1)$, compute the  IS weights}
\begin{equation}
w_1^i=\frac{g_\theta(y_1|x_1^i)f_\theta(x_1^i)\chi_2(x_1^i;\hat c_2)}{q_1(x_1^i;\hat c_1)},
\end{equation}
{\it store $\bar w_1= \sum_{i=1}^N  w_1^i/N$, and  compute normalized weights $W_1^i=w_1^i/(\sum_{l=1}^N w_1^l)$ . }
\item[-] {\it If resampling, sample $\bar x_1^i\sim \sum_{i=1}^N W_1^i\delta_{x_1^i}(dx_1)$ and set the IS weights to
                 $W_1^i=1/N$, otherwise set $\bar x_1^i=x_1^i$}.
\end{itemize}
\item[] {\it For $t=2,...,T$:}
\begin{itemize}
\item[-] {\it Sample $x_t^i\sim q_t(x_t|\bar x_{t-1}^i,\hat c_t)$ and set $x_{1:t}^i=(x_t^i,\bar x_{1:t-1}^i)$;}
\item[-] {\it compute the IS weights}
\begin{equation}
        w_t^i= W_{t-1}^i  \frac{g_\theta(y_t|x_t^i)f_\theta(x_t^i|x_{t-1}^i)\chi_{t+1}(x_{t}^i,\hat c_{t+1}) }{k_t(x_{t}^i,x_{t-1}^i;\hat c_t)},
\end{equation}
{\it store $\bar w_t= \sum_{i=1}^N  w_t^i$, and  compute normalized weights $W_t^i=w_t^i/(\sum_{l=1}^N w_t^l)$. }
\item[-] {\it If resampling, sample $\bar x_{1:t}^i\sim \sum_{i=1}^N W_t^i\delta_{x_{1:t}^i}(dx_{1:t})$ and set the IS weights to
 $W_t^i=1/N$, otherwise set $\bar x_{1:t}^i=x_{1:t}^i$.}
\end{itemize}
\item[(iii)] {\it If required, compute the SMC likelihood estimate according to  (\ref{eq:smc-app-integrating const}):}
\begin{equation}\label{eq:PEIS-likelihood}
\hat z_T = \hat p_\theta(y_{1:T})=\chi_1(\hat c_1) \displaystyle \prod_{t=1}^T \bar w_t.
\end{equation}
\end{itemize}
\end{itemize}

\doublespacing

As discussed above, PEIS differs from standard SMCs such as BPF and APF by its forward-looking design incorporating  information in the data $y_{t:T}$ into the targets $\pi_t(x_{1:t})$ and  sampling densities  $q_t$. A further difference is that in PEIS the parameters  of the targets and sampling densities are, in contrast to standard SMCs, random variables  since they  depend via the optimal EIS auxiliary parameters $\hat c$ on the EIS CRNs $\bar u_{1:T}$
(see step (i) of the PEIS algorithm). However, the  data $y_{1:T}$ is fixed as it is also the case for the EIS parameters $\hat c$ when it comes to generating the SMC particles $\bar x_{1:T}$ and ancestor indices
$\bar a_{1:T-1}$ in step (ii) of the PEIS algorithm. With fixed $y_{1:T}$ and $\hat c$ neither the forward-looking design nor the dependence of $\hat c$
on the CRNs $\bar u_{1:T}$ changes the inherent structure in the joint distribution  of the SMC particles $\bar x_{1:T}$ and  ancestor indices $\bar a_{1:T-1}$ defined by a standard SMC or its conditional SMC counterpart required by PG. It follows that conditional on $\bar u_{1:T}$ the PEIS defines a valid  SMC representing a special case of the generic  SMC algorithm  for the PMCMC framework of Andrieu et al.~(2010). Also it produces by $\hat p_\theta(y_{1:T})$ in  (\ref{eq:PEIS-likelihood})  unbiased estimates of the likelihood $p_\theta(y_{1:T})$ (Scharth and Kohn, 2016).

\subsubsection*{4.1.3 PG with PEIS }
Since  PEIS  defines conditional on the EIS CRNs a valid SMC, we can suggest to use  it in the PG algorithm in Section 3.1.
For the validity of the resulting PG-PEIS procedure, however, it is critical that we include those CRNs into the PG Markov kernel. This can be easily implemented by  drawing in each PG sweep $j$ a new set of  CRNs  $\bar u_{1:T}^{(j)}$ to produce via the initial EIS regressions (\ref{eq:EIS-LS}) the auxiliary parameters $\hat c^{(j)}$  for the PEIS version of the conditional SMC step targeting  $p_{\theta^{(j)}}(x_{1:T}|y_{1:T})$.
It follows that such a PG-PEIS procedure defines a Markov kernel including updates for the CRNs $\bar u_{1:T}$ with a target density obtained as the following straightforward extension of the PG target in  (\ref{extended PG target}):
\begin{equation}\label{extended PG-PEIS target}
\tilde p(\theta,k, \bar x_{1:T}, \bar a_{1:T-1},\bar u_{1:T} )=\frac{1}{N^T}p(\theta,x_{1:T}^k|y_{1:T})p(\bar u_{1:T})
\tilde p( \bar x_{1:T}^{\backslash b_{1:T}^k}, \bar a_{1:T-1}^{\backslash b_{1:T}^k}| x_{1:T}^k,b_{1:T}^k,\theta,\bar u_{1:T}),
\end{equation}
where $p(\bar u_{1:T})$ is the canonical density of the EIS CRNs. The function $\tilde p$ on the r.h.s represents the density of the random variables generated by the conditional PEIS given $\theta, x_{1:T}^k, b_{1:T}^k, \hat c$, where we can replace $\bar u_{1:T}$ by  $\hat c$ since  $\hat c$ given $\theta$ obtains via the EIS regressions (\ref{eq:EIS-LS}) as a deterministic function of $\bar u_{1:T}$. Note that the updating step for $\bar u_{1:T}$ consists of drawing from its marginal density $p(\bar u_{1:T})$, which implies that $\bar u_{1:T}$ is marginalized out in the individual Gibbs sweeps. However, this marginalization leaves the extended target density $\tilde p(\theta,k, \bar x_{1:T}, \bar a_{1:T-1},\bar u_{1:T} )$ invariant (Liu, 1994). Moreover, the extended target (\ref{extended PG-PEIS target}) includes in the same way as the PG target in  (\ref{extended PG target}) the original target density   $p(\theta,x_{1:T}^k|y_{1:T})$ as a marginal. It follows that the augmented  Markov kernel defined by the PG-PEIS leaves $p(\theta,x_{1:T}^k|y_{1:T})$ invariant.

As discussed above, the PEIS when implemented with SMC-resampling steps in every period will suffer, as any SMC, from the SMC-path degeneracy  causing the poor mixing of the baseline PG. Hence, we can not expect that such a brute-force PG-PEIS implementation will satisfactorily address the poor mixing problem of PG. However, since PEIS is designed to globally minimize the variance of the SMC IS weights, we expect that it suffices to resample only at a few periods,  which substantially reduces the path degeneracy.
This  motivates the implementation of the baseline PG using the PEIS with sparse resampling at a few predetermined time periods  (PG-PEIS-sparse). Moreover, the PG-PEIS framework offers the possibility to substantially improve the PG extensions which we discuss next.

\subsection*{4.2 Particle Gibbs with ancestor sampling (PGAS)}
In order to address the poor mixing of the baseline PG,  Lindsten et al.~(2014) developed the  PGAS. It exploits the fact that it suffices to suppress the degeneration of the particle system towards the retained conditional SMC reference trajectory $x_{1:T}'$ and not the SMC-path degeneracy per se to improve the mixing. Based on this insight, the basic idea of the PGAS is to break this reference trajectory into pieces, so that the particle system tends to degenerate to something different than the reference trajectory.

In particular, the PGAS augments each period-$t$ conditional-SMC resampling step  by randomly selecting from the set $\{x_{1:t-1}^i\}_{i=1}^N$ (including the reference particle $x_{1:t-1}'=x_{1:t-1}^1$)  one ancestor particle which is used to assign a potentially  new $x_{1:t-1}$-history  to the partial frozen path $x_{t:T}'$. This produces a  concatenated full path $[x_{1:t-1}^i,x_{t:T}']$, and  the corresponding (non-normalized) weight  for selecting $x_{1:t-1}^i$ as the new ancestor for $x_{t:T}'$    is given by
\begin{equation}\label{eq:PGAS-weights}
\tilde w_{t-1|T}^i=w_{t-1}^i\;\frac{\gamma_T\left([x_{1:t-1}^i,x_{t:T}']\right)}{\gamma_{t-1}(x_{1:t-1}^i)}.
\end{equation}
In Bayesian terms, the components of those  ancestor sampling (AS)  weights for the reference particle  are the prior probability of the ancestor particle $x_{1:t-1}^i$ given by the `standard' SMC-IS weights $w_{t-1}^i$ and the likelihood that the partial path $x_{t:T}'$ originated from  $x_{1:t-1}^i$  represented by the ratio of the targets $\gamma_T/\gamma_{t-1}$.

As shown by Lindsten et al.~(2014, Theorem 1), the invariance property of the baseline PG is not violated by  this additional AS step.
However, since this AS step sequentially assigns  in each period a  potentially new ancestor to $x_{t:T}'$, it will produce a reference path $x_{1:T}'$ which tends to differ from the other (degenerated) conditional SMC paths $\{x_{1:T}^i\}_{i=2}^N$. Thus, while not preventing the particle system to degenerate, the PGAS  typically improves the mixing of the baseline PG.
Furthermore, the larger the sum of the AS weights $\tilde w_{t-1|T}^i$ for the set of potential new ancestors $\{x_{1:t-1}^i\}_{i=2}^N$ relative to that for the old ancestor $x_{1:t-1}'$, the higher the probability that the old history is replaced a new one. Hence, by increasing the relative AS weights for the new ancestors, we can  increase the potential diversity  of the resulting PGAS reference path so as to improve the mixing of the PGAS trajectories $x_{1:T}^{(j)}$ under $p_\theta(x_{1:T}|y_{1:T})$.

In Lindsten et al.~(2014), the PGAS is implemented by relying upon the BPF  (PGAS-BPF), which uses $\pi_t(x_{1:t})\propto p_\theta(x_{1:t},y_{1:t})$, as given in (\ref{eq:gamma_t}),
together with $q_t(x_t|x_{1:t-1})\equiv f_\theta (x_t|x_{t-1})$, so that according to
(\ref{eq:SMC-is-weights})
the prior weights for $x_{t-1}^i$ are $w_{t-1}^i = W_{t-2}^i g_\theta(y_{t-1}|x_{t-1}^i)$. The resulting AS weights
are given by
\begin{equation}\label{eq:BPF-AS-weights}
\tilde w_{t-1|T}^i=W_{t-2}^i  g_\theta(y_{t-1}|x_{t-1}^i)p_\theta(x_{t:T}',y_{t:T}|x_{t-1}^i)\propto W_{t-2}^i g_\theta(y_{t-1}|x_{t-1}^i)f_\theta(x_t'|x_{t-1}^i),
\end{equation}
where $f_\theta(x_t'|x_{t-1}^i)$ represents the likelihood  for $x_{t-1}^i$. As to the probability of replacing the old by a new history, we note that since
the old ancestor $x_{t-1}'$ has produced $x_t'$ using the BPF IS density $f_\theta(x_t|x_{t-1}')$, the position of $x_{t-1}'$ is adapted to $x_t'$ which locates the likelihood.  As a draw from the BPF approximation to the smoothing density  $p_\theta(x_{t-1}|y_{1:T})$ the $x_{t-1}'$-position  is also adapted to $y_{t-1}$ which allocates the probability mass of the prior weights.
In contrast, the potential new ancestors $\{x_{t-1}^i\}_{i=2}^N$ weighted by $\{W_{t-2}^i\}_{i=2}^N$ represent draws  from the BPF approximation to the predictive  density $p_\theta(x_{t-1}|y_{1:t-2})$ so that they are positioned  blindly  w.r.t.~$y_{t-1}$ and $x_t'$.
This implies that if the measurement density $g_\theta$  and the transition  $f_\theta$ are (as functions in $x_{t-1}$)  tight and have domains which are poorly aligned with that of the predictive density  $p_\theta(x_{t-1}|y_{1:t-2})$, then the resulting prior weights $g_\theta(y_{t-1}|x_{t-1}^i)$ and likelihood values $f_\theta(x_t'|x_{t-1}^i)$ for $x_{t-1}'$ can be substantially larger than those for $\{x_{t-1}^i\}_{i=2}^N$.
This would produce an  AS weight for the old ancestor which is large relative to those for the potential new ancestors. Hence, in applications prone to such a poor alignment of the predictive densities the efficacy of the PGAS-BPF to improve the mixing of the baseline PG can be expected to be limited.
These are especially applications with a tight measurement density coupled with measurement outliers.

The  globally fully adapted SMC targeted by PEIS  uses according to (16) and (19) $\pi_t(x_{1:t})\propto$\linebreak $p_\theta(x_{1:t},y_{1:t})p_\theta(y_{t+1:T}|x_t)$ and $q_t(x_t|x_{1:t-1})\equiv p_\theta (x_t|x_{t-1},y_{t:T})$ so that the prior weights are $w_{t-1}^i\propto 1$. Hence, the resulting AS weights only depend on the likelihood and become
\begin{align}\label{eq:opt-AS-weights}
\tilde w_{t-1|T}^i&\propto 1\cdot \frac{p_\theta(y_{t:T},x_{t:T}'|x_{t-1}^i)}{p_\theta(y_{t:T}|x_{t-1}^i)} \propto   1\cdot  p_\theta(x_{t}'|x_{t-1}^i, y_{t:T}).
\end{align}
Here, the potential ancestors  $\{x_{t-1}^i\}_{i=2}^N$ and $x_{t-1}'$  all represent  draws from  $p_\theta(x_{t-1}|y_{1:T})$ so that their positions are  equally well adapted to $y_{t-1:T}$ which is why they all have  the same prior weights $w_{t-1}^i$. Moreover, as the potential new ancestors $\{x_{t-1}^i\}_{i=2}^N$ are draws from the same distribution as the old ancestor $x_{t-1}'$, which has produced $x_t'$ using the fully adapted IS density $p_\theta(x_{t}|x_{t-1}', y_{t:T})$,
it is also the case that the potential new ancestors are placed in the state-space system such that their expected likelihood values $p_\theta(x_{t}'|x_{t-1}^i, y_{t:T})$ is as large as possible relative to that of the old ancestor.
Under the PEIS designed to provide the best possible approximation to the  globally fully adapted SMC, the constant prior weights in (\ref{eq:opt-AS-weights}) are replaced by their PEIS approximation in (\ref{PEIS-is-weights}) (shifted by one period) and the optimal IS density defining the likelihood in (\ref{eq:opt-AS-weights}) by its  PEIS approximation in (\ref{eq:IS-density-EIS-1}). Hence, PEIS is designed so as to produce  a high potential diversity of the reference particle generated by the additional AS step. As a result, we expect to improve the mixing of the PGAS paths for $x_{1:T}$ obtained under local procedures like the BPF by relying upon the global PEIS (PGAS-PEIS). This improvement can be expected to be especially large in applications prone to a poor alignment of the predictive densities.

As to the validity of the PGAS-PEIS, we note that Theorem 1 in Lindsten et al.~(2014) establishing the invariance property of the PGAS kernel is not affected by the use of PEIS.
In fact, when using the PEIS this invariance property of the PGAS holds true  conditional on the EIS CRNs $\bar u_{1:T}$  so  that we can augment the PGAS kernel  in line with with the augmentation discussed in Section 4.1.3  to include the (marginal) update steps for $\bar u_{1:T}$  which preserves the invariance.

\subsection*{4.3 Particle Gibbs with an additional MH step (PGMH)}
The PGMH proposed by Holenstein (2009, Algorithm 3.6) in order to address the poor-mixing problem of the baseline PG bypasses the SMC-path degeneracy by using an additional particle-MH step  proposing in each iteration step $j$ a completely new SMC path denoted by $x_{1:T}^*$. This new path is MH-compared with the old path $x_{1:T}^{(j-1)}$ based upon the (conditional) SMC estimates of their respective marginal likelihood. The PGMH algorithm is given by:

\onehalfspacing
\begin{itemize}
\item[]{\bf \underline {PGMH algorithm}}
\begin{itemize}
\item[(i)] {\it Initialization $(j=0)$: Set arbitrary $\theta^{(0)}$, run an SMC targeting $p_{\theta^{(0)}}(x_{1:T}|y_{1:T})$,
       and sample $x_{1:T}^{(0)}\sim \hat p_{\theta^{(0)}}(x_{1:T}|y_{1:T})$.}
\item[(ii)] {\it For iteration $j\geq 1$:}
\begin{itemize}
\item[-]{\it  sample  $\theta^{(j)}\sim p(\theta| x_{1:T}^{(j-1)},y_{1:T} )$ },
\item[-]{\it  run a conditional SMC targeting  $p_{\theta^{(j)}}(x_{1:T}|y_{1:T})$  conditional on $x_{1:T}^{(j-1)}$,
       and  compute the likelihood estimate $\hat p_{\theta^{(j)}}(y_{1:T})$,}
\item[-] {\it  run an SMC targeting   $p_{\theta^{(j)}}(x_{1:T}|y_{1:T})$, sample $x_{1:T}^{*}\sim \hat p_{\theta^{(j)}}(x_{1:T}|y_{1:T})$, and compute the likelihood estimate $\hat p_{\theta^{(j)}}^*(y_{1:T})$,}
\item[-]  {\it  with probability}
\begin{equation}\label{eq:accept-prob-PGMH}
      1 \wedge\frac{\hat p_{\theta^{(j)}}^*(y_{1:T})}{\hat p_{\theta^{(j)}}(y_{1:T})}
\end{equation}
      {\it  set $x_{1:T}^{(j)}= x_{1:T}^{*}$, otherwise $x_{1:T}^{(j)}= x_{1:T}^{(j-1)}$. }
\end{itemize}
\end{itemize}
\end{itemize}

\doublespacing

The efficacy of this PGMH algorithm to improve the mixing of the baseline PG critically depends on the numerical precision of the (conditional) SMC estimates for the marginal likelihood $p_\theta(y_{1:T})$ defining the acceptance rate of the additional particle MH-step as given in  (\ref{eq:accept-prob-PGMH}). In particular, if the SMC delivers noisy estimates for $p_\theta(y_{1:T})$ the MH updates for $x_{1:T}$ can get stuck for many iterations leading to very poor mixing. Hence, efficient PGMH implementations are those for which the SMC marginal likelihood estimates  have a small variance. Since, as discussed in Section 4.1,  PEIS aims at  producing very precise SMC estimates, we expect a high efficacy of the PGMH in improving the mixing of the baseline PG by relying upon PEIS estimates for
$p_\theta(y_{1:T})$ as given in   (\ref{eq:PEIS-likelihood}) (PGMH-PEIS).

Note that this PGMH-PEIS  implemented with redrawing in each sweep $j$ the EIS-CRNs $\bar u_{1:T}$ has the desired invariant distribution, which follows  directly from the target-augmentation principle discussed in Section 4.1.3 and the fact that the PEIS  estimates of the marginal likelihoods in (\ref{eq:accept-prob-PGMH}) are unbiased (Scharth and Kohn, 2016).

\section*{5. Univariate Applications}
\subsection*{5.1 Example Models}
The first example is a standard stochastic volatility (SV) model for the volatility of financial returns (see, e.g., Ghysels et al., 1996). It has the form
\begin{eqnarray}\label{eq:SV-measure}
y_t&=&\beta \exp\{x_t/2\}\eta_t,\qquad \eta_t\sim \mbox{i.i.d.}N(0,1),\\
x_t&=&\delta x_{t-1} + \nu \epsilon_t,\qquad \epsilon_t\sim \mbox{i.i.d.}N(0,1),\label{eq:SV-trans}
\end{eqnarray}
where $y_t$ is the asset return observed at period $t$, $x_t$ is the latent log volatility and $\theta=(\beta,\delta,\nu)'$. The innovations $\epsilon_t$ and $\eta_t$ are mutually independent.

The second example is  a time-discretized version of the shifted Cox-Ingersoll-Ross (1985) diffusion (CIR) for the daily shadow rate as proposed by  Gorovoi and Linetsky (2004). It is based on Black's (1985) approach of interest rates as options which assumes that there is a shadow interest rate that can become negative, while the nominal rate is the positive part of the shadow rate. In order to account for microstructure noise, which is to be expected for interest rate data at the daily frequency, the shifted CIR specification is extended  to include a noise component  (A\"{\i}t-Sahalia, 1999). The resulting model for the nominal interest rate $y_t$  observed at day $t$ with $x_t\in(\kappa,\infty)$ the latent  shadow rate, is described as
\begin{eqnarray}\label{eq:CIR-measure}
y_t&=&\max(x_t,0) + \sigma_y\eta_t,\qquad \eta_t\sim \mbox{i.i.d.}N(0,1),\\
x_t&=&x_{t-1} +\Delta(\alpha-\beta x_{t-1}) +\sigma_x \sqrt{x_{t-1}-\kappa}\sqrt{\Delta}\epsilon_t,\qquad \epsilon_t\sim \mbox{i.i.d.}N(0,1),\label{eq:CIR-trans}
\end{eqnarray}
where $\epsilon_t$ and $\eta_t$ are independent, $\Delta=1/252$, and $\kappa<0$ is the shift factor. Following Gorovoi and Linetsky (2004), we set $\kappa=-0.05$. The parameters are $\theta=(\alpha,\beta,\sigma_x,\sigma_y)'$.

The data we use for the SV model are daily log returns of the S\&P 500 stock index from Oct.~1, 1999 to Sep.~30, 2009, with a sample size of $T=2515$.
For the CIR model the data consists of daily 3-month US-Treasury bill rates on the secondary market from Jan.~3, 2000 to Feb.~1, 2018, with $T=4525$. See Figure 1 for time-series plots of the data.

In the SV return data the parameter estimates imply a measurement density which is fairly uninformative about the states representing a scenario where standard SMCs typically exhibit satisfactory performance. In contrast, the measurement density for the fitted CIR model is very informative about the states leading to a large sensitivity of standard SMCs to measurement outliers with potential adverse effects on the efficiency of  PG implementations. Such outliers are frequently observed in the interest rate data.

In the online appendix we provide a full description of the (P)EIS implementation used for the two example models. The SV example involves a Gaussian state transition ($f_\theta$) suggesting to select for the EIS implementation  Gaussian  EIS kernels $k_t$ obtained as a parametric extensions of $f_\theta$. Under the CIR model the product of the measurement and transition density ($g_\theta f_\theta$) defines a kernel of a mixture of two truncated Gaussian densities for $x_t$ given $x_{t-1}$ so that we can construct $k_t$ as  parametric extensions of $g_\theta f_\theta$. In both cases the  EIS LS regressions (\ref{eq:EIS-LS}) take the form of linear low-dimensional LS problems.  We run those regressions using $R=15$ trajectories $\{x_{1:T}^i\}_{i=1}^R$  and repeat them for 4 fixed-point iterations. The $R^2$ we find in the final iteration is typically larger than 0.99, which indicates that the resulting PEIS densities defined in (\ref{eq:IS-density-EIS-1}) are  well  globally adapted to the SMC target $p_\theta(x_{1:T}|y_{1:T})$.

\subsection*{5.2 Results}
Here we present simulation experiments using  the SV and CIR model to compare the following 8 PG schemes:
The baseline PG based on the BPF (PG-BPF),  PEIS (PG-PEIS) and  PEIS with sparse resampling (PG-PEIS-sparse), then the PGAS combined with the BPF (PGAS-BPF) and  PEIS (PGAS-PEIS) and, finally, the PGMH using the BPF (PGMH-BPF),  PEIS (PGMH-PEIS) and PEIS-sparse  (PGMH-PEIS-sparse). We use multinomial resampling for the SMC resampling steps. For the PEIS-sparse the resampling is conducted only every 500 periods. The  PG methods were all implemented  in the interpreted  language MATLAB, making computing times comparable.

For all the experiments  we use the real data sets described in Section 5.1. The corresponding maximum likelihood (ML) estimates based on  EIS evaluations of the likelihood are  $(\beta,\delta,\nu)$ = (1.065, 0.992, 0.122) for the SV model and $(\alpha,\beta,\sigma_x,\sigma_y)$ = (0.0013, 0.2179, 0.0287, 9.8e-5) for the CIR model.

\subsubsection*{5.2.1 Mixing of Particle Gibbs for fixed parameters}
The first experiment is designed to analyze the mixing of the PG algorithms w.r.t.~the states under their joint posterior $p_\theta(x_{1:T}|y_{1:T})$ for a fixed value of the parameters $\theta$. Throughout this experiment we set the parameters equal to their ML estimates and generate samples from this density using the PG algorithms, which are all implemented with two different numbers of particles, $N=30$ and $N=1000$. All methods are simulated for 1100 iterations, where the first 100 burn-in iterations are discarded.

In order to compare the mixing, we  compute the update rate for each $x_t$ $(t=1,...,T)$ which is defined as the proportion of PG iterations where the value for $x_t$ has changed. The update rates for the 8 PG algorithms  plotted against time $t$ are provided in Figure 2 for the SV model and in Figure 3 for the CIR model. They reveal  that in both example models  the update rate for the baseline PG-BPF for $N=30$ as well as $N=1000$  rapidly decreases for an increasing distance of $t$ to the final period $T$. These poor update rates  reflect the  SMC path degeneracy causing, as discussed in Section 3.2, the state trajectory $x_{1:T}^{(j)}$ at PG iteration step $j$ to coalesce with the previous trajectory $x_{1:T}^{(j-1)}$.
That this poor mixing problem cannot be addressed satisfactorily by replacing the locally designed BPF by an SMC which is well globally adapted  is evidenced by the update rates of the PG-PEIS: Even if they increase relative to the PG-BPF they fall in both models, even with $N=1000$ particles, below 20\%  for the states of the first 500  periods. This is an illustration of the `unavoidable' SMC path degeneracy which we would obtain under a fully optimal SMC when resampling is performed every period.   The update rates for the PG-PEIS-sparse  remaining above 50\% across all periods  show  that, as expected,  sparse resampling  greatly  improves the mixing of the PG-PEIS by reducing the path degeneracy effects.  For $N=1000$ they are nearly completely eliminated indicating  a very small variance of the PEIS-sparse SMC weights used to resample every 500 periods.

The comparison of the baseline PG-BPF with the PGAS-BPF shows that the additional AS step also increases significantly the average probability of updating $x_t$ across all periods. However, for the CIR  model, in particular, this probability drops dramatically in many periods, indicating that in these periods  very few particles tend to keep all the weights across the PG iterations. As discussed in Section 4.2, this stems from the model's tight measurement density which makes the PGAS-BPF particularly vulnerable  to  measurement outliers as they produce a very small probability of replacing the old by a new history.
This effect appears to be  less acute for the SV model reflecting the  fact that its measurement distribution is not very sensitive to the state. When combined with PEIS, the PGAS with as little as $N=30$ particles produce update rates which are uniformly above 95\% for both, the SV model and the CIR model, indicating a close to  ideal   and robust mixing of the PGAS.

Turning to the PG augmented by an additional MH move, we also find in both example models a substantial improvement in the mixing when replacing the BPF by PEIS or PEIS-sparse. Those improvements reflect the fact that, as discussed in Section 4.3, PEIS(-sparse) produce numerically  far more accurate SMC estimates of the full marginal likelihood than the BPF.
However, even with $N=1000$ there is still some noise in the PEIS(-sparse) full likelihood estimates, especially in the challenging CIR application.
This explains why for $N=1000$ the additional MH move
in the PGMH-PEIS(-sparse)   based on those likelihood estimates  reduces the update rates relative to those of the direct PG-PEIS-sparse which itself
is  for $N=1000$ hardly hindered by any weight- and path-degeneracy.

For a further comparison of the PG methods, we compute the effective sample size
of the posterior samples for the state variable $x_t$ at each  period $t$. It is defined as $\mbox{ESS}=M\cdot \left[ 1+2\textstyle \sum_{j=1}^J\gamma(j)\right]^{-1}$,
where $M$ is the size of  the posterior sample, and $\sum_j\gamma(j)$  the sum of the $J$ monotone sample autocorrelations (Geyer, 1992).
The interpretation  is that the $M$ PG draws lead to the same precision as a hypothetical i.i.d.~sample from the posterior of size ESS, so that large values for ESS are preferable. We consider the minimum, median and maximum ESS over the $T$ sampled state variables. These ESS values are computed  for 10 independent complete PG runs from which we take the corresponding averages. In order to account for different computing times, we also compute the (average) minimum ESS standardized by the Central Processor Unit (CPU) time required to run a PG algorithm. It measures the time it takes to obtain one effective draw of the {\sl complete} $x_{1:T}$-trajectory from its posterior. The ESS results are reported in Table 1 for the SV model and in Table 2 for the CIR model.

The results for both models show that, for a given number of particles $N$, the PEIS(-sparse) substantially increases the median and the minimum ESS of the baseline PG, PGAS and PGMH relative to their corresponding BPF counterpart. The largest  effective sample 
per hour computing time is produced by  the PG-PEIS-sparse  with $N=1000$ for the SV model, and by
the PGAS-PEIS with $N=30$ for the CIR model. This illustrates that the improvements of the PG approximations to  $p_\theta(x_{1:T}|y_{1:T})$ gained by the global PEIS outweigh its   additional computational costs relative to the locally designed BPF.

\subsubsection*{5.2.2 Full Bayesian  analysis}
Here, we compare the performance of the PG algorithms for a full Bayesian analysis of the two example models. For the parameters of both models we select fairly uninformative priors (for details of the prior selection, see the online appendix). In light of the severe mixing problems  of the PG-BPF, PG-PEIS and PGMH-BPF documented in the previous section, the remainder investigation focuses on the efficiency of the PG-PEIS-sparse, PGAS-BPF, PGAS-PEIS, PGMH-PEIS and PGMH-PEIS-sparse.
For all of those five methods we use throughout 50,000 PG iterations  where the first 10,000 burn-in iterations are discarded.

The Bayesian posterior results for the five PG procedures, each based on $N=30$ particles, are summarized in Table 3 for the SV model and in Table 4 for the CIR model. They report the  PG posterior mean and standard deviation of the parameters together with the ESS and  ESS standardized by computing time.  All reported statistics  are sample averages which are computed from 10 independent replications obtained by running each of the PG algorithms under 10 different seeds. The tables also provide the results  for the `ideal' Gibbs sampler, i.e., the sampler which simulates $x_{1:T}$  directly  from the true posterior $p_\theta(x_{1:T}|y_{1:T})$. This fictitious Gibbs sampler is approximated   by the PGAS-PEIS implemented with $N=10,000$ particles. Since the PG algorithms can be seen as MC approximations of the ideal Gibbs sampler, the latter provides a natural benchmark for the mixing performance of the former (Lindsten et al.~2014).

From the results for the SV model in Table 3   we see that  all five PG algorithms produce MC estimates of the posterior means which  are close to those of the ideal Gibbs sampler and the corresponding ML estimates. The ESS values indicate that  replacing the BPF by  PEIS improves, as expected from the results in Section 5.2.1, the mixing of the PGAS for the parameters, and shifts the ESS values closer to those of the ideal Gibbs sampler. The remaining  PG-schemes based upon PEIS or PEIS-sparse also show a satisfactory mixing relative to the ideal Gibbs. Most critical for a posterior Gibbs  analysis of the parameters $\theta$ appears to be the  parameter $\beta$, which has among all parameters and across all PG procedures the smallest ESS. Hence, the mixing of the sampled $\beta$'s sets the limit w.r.t.~the amount of  effective draws for $\theta$ which can be generated for a given number of Gibbs iterations or a fixed computing time.  In terms of the largest  minimum ESS of the sampled parameters per hour computing time, the PG-PEIS-sparse and PGAS-BPF show the best performance. Both produce per hour 7  effective draws from the  parameters' marginal posterior  $p(\theta|y_{1:T})$.

Turning to the results for the CIR model in Table 4, we first note that the MC estimates for the parameters' posterior mean  obtained from PG procedures based on PEIS are all  virtually the same as those of the ideal Gibbs and their ML counterparts. For the PGAS based on BPF, however, the posterior parameter estimates differ from those benchmarks especially that for $\sigma_y$.
These  biases are consistent with the results of Section 5.2.1, showing that in situations involving tight measurement densities coupled with outliers the PGAS-BPF has severe problems to fully explore the domain of the states under $p_\theta(x_{1:T}|y_{1:T})$. That the measurement density  is fairly tight is indicated by  the tiny value of the estimates for $\sigma_y$.
In contrast to the PGAS-BPF, the PG procedures based on PEIS ensure even in this challenging scenario  a
fast and reliable exploration of $p_\theta(x_{1:T}|y_{1:T})$ and lead to  accurate posterior estimates for the parameters. Note also that  the ESS values in Table 4 indicate that the mixing rate of all PG procedures using PEIS is very close to that of the ideal Gibbs sampler. (Since the PGAS-BPF parameter draws 	 apparently   fail to appropriately represent the posterior $p(\theta|y_{1:T})$ we refrain from reporting their  ESS values.)

In the online appendix we provide the results of a robustness analysis of the PGAS implementations for the Bayesian analysis of the SV and CIR model w.r.t.~the number of SMC particles $N$. It also provides additional results for  the SV model comparing the PG procedures with the wildly used Gibbs approach of Kim et al.~(1998) based on an auxiliary mixture sampler specifically tailored to the analysis of SV models.

\section*{6. Multivariate Application}
\subsection*{6.1 Model}
We now turn to a multivariate example where we consider the multivariate SV model of Tsay (2010, Section 12.7.2) adapted to the modelling of the realized covariance matrices observed for a set of $r$ asset returns (for a discussion of realized covariances, see Barndorff-Nielsen and Shephard, 2004).
For the $r\times r$ realized covariance matrix $Y_t$ we assume an inverted Wishart distribution  with density
\begin{equation}\label{eq:IW}
g_\theta(Y_t|\Sigma_t)=\frac{|\Sigma_t|^{\nu/2} |Y_t|^{-(\nu+r+1)/2}\exp\{-(1/2)\mbox{tr}(\Sigma_t Y_t^{-1})\}}{2^{\nu r/2}\pi^{r(r-1)/4}\prod_{s=1}^r\Gamma([\nu+1-s]/2)},
\end{equation}
where $\Sigma_t$ is the period-$t$ positive definite $r\times r$ scale matrix and $\nu>r+1$ the degrees of freedom so that $E(Y_t|\Sigma_t)=\Sigma_t/(\nu-r-1)$ (Anderson, 1984).
The time-varying scale matrix  $\Sigma_t$ directing the  conditional mean of $Y_t$  is Cholesky-decomposed and is taken to depend upon a latent Gaussian autoregressive state vector $x_t=(x_{1t},...,x_{rt})$ in the form
\begin{align}\label{eq:IWscale1}
\Sigma_t  & = HD_tH',\quad D_t=\mbox{diag}\{\exp(x_{1t}),\ldots,\exp(x_{rt})\},\\
\label{eq:IWscale2} x_{s t} & =\mu_s+\delta_s(x_{s t-1}-\mu_s)+\sigma_s\epsilon_{s t},\quad \epsilon_{s t}\sim \mbox{i.i.d.}N(0,1),\quad s=1,\ldots,r.
\end{align}
The matrix $H$ is a  lower-triangular  $r\times r$ parameter matrix with unit diagonal elements. Its column vectors denoted by $h_s$ $(s=1,\ldots, r)$ have the form $h_s=(0,\ldots,0,1,\tilde h_s')'$, where $\tilde h_s=(\tilde h_{s,1},\ldots,\tilde h_{s,r-s})'$ is the lower subvector of $h_s$ consisting of unrestricted parameters to be estimated. Thus we have $\theta=(\nu,\mu_1,\delta_1,\sigma_1,\ldots,\mu_r,\delta_r,\sigma_r,\tilde h_1',\ldots,\tilde h_{r-1}')'$.

Under the assumed specification for $\Sigma_t$ the measurement density in  (\ref{eq:IW}) factorizes as a function in $x_t$ into
\begin{equation}\label{eq:IWmeasurement}
g_\theta (Y_t|\Sigma_t)=\prod_{s=1}^r g_\theta(\tilde y_{s t}|x_{s t}),\quad\mbox{with}\quad g_\theta(\tilde y_{s t}|x_{s t})\propto \exp\Big\{\frac{\nu}{2}x_{s t}-\frac{1}{2}\tilde y_{s t} \exp(x_{s t})\Big\},
\end{equation}
where $\tilde y_{s t}=h_s'Y_t^{-1}h_s$. It follows that the $r$ state processes $x_{1:T}=(x_{1,1:T},\ldots,x_{r,1:T})$ are mutually independent under their joint conditional posterior 
given by
\begin{align}
p_\theta(x_{1:T}|Y_{1:T})&=\prod_{s=1}^r p_\theta(x_{s,1:T}|Y_{1:T}),\\
 \mbox{with}  &\qquad p_\theta(x_{s,1:T}|Y_{1:T})\propto \Bigg[\prod_{t=2}^T g_\theta(\tilde y_{s t}|x_{s t}) f_\theta(x_{s t}|x_{s t-1})\Bigg]
                         g_\theta(\tilde y_{s 1}|x_{s 1}) f_\theta(x_{s 1}),
\end{align}
where $f_\theta$ represent the Gaussian transitions defined in  (\ref{eq:IWscale2}). This independence allows us to parallelize in the PG procedures the conditional SMC step for $p_\theta(x_{1:T}|Y_{1:T})$ by running separately for each  $s=1,\ldots,r$ a conditional SMC  targeting $p_\theta(x_{s,1:T}|Y_{1:T})$. The corresponding parallelized conditional PEIS  requires only minor modifications of the PEIS implementation for the univariate SV model in Section 5.2 (see the online appendix).

\subsection*{6.2 Results}
The inverted Wishart SSM is applied to $T=2514$ daily realized covariance matrices observed  for $r=5$  stocks  (American Express, Citigroup, General Electric, Home Depot, and IBM). The data spans Jan.~1, 2000 to Dec.~31, 2009 (for a detailed discussion of this data, see Golosnoy et al., 2012).

 We use the PGAS-BPF and PGAS-PEIS for a full posterior MCMC analysis assuming  independent conjugate priors for the $\mu_s$'s (Gaussian),  $\sigma_s^2$'s (inverse Gaussian) and $\tilde h_s$'s (Gaussian), and uniform priors for the $\delta_s$'s and $\nu$ (for details of the prior selection, see the online appendix). Both PGAS methods are simulated  for 15,000 iterations discarding  the first 5,000 samples as burn-in.  All computations are performed using MATLAB on a 3.1 GHz Intel Core i5 processor. The parallel updating of the state processes $x_{1:T}$  are distributed  on 4 cores using the MATLAB function {\tt parfor}. As for the univariate applications, we use for the EIS LS regressions $R=15$ trajectories and repeat them for 4 fixed-point  iterations.

In Figure  4 we display the update frequencies for the $r=5$  state processes under the PGAS-BPF with $N=30$ and $N=1000$ particles and the PGAS-PEIS with $N=30$. They reveal  that the PGAS-BPF with $N=30$ suffer from update rates which fall  substantially below 50\% for many periods leading to $x_{s t}^{(j)}$-chains  which are stuck for many iterations. Even an increase of the particle number to $N=1000$ does not fully prevent occasional low update rates. In sharp contrast, the PGAS-PEIS produces with as little as $N=30$ particles update frequencies which are uniformly  close to the maximum ($1-1/N\simeq 0.97$) leading to a  mixing for all the $T \cdot r = 12570$ state variables, which is close to that of the `ideal' Gibbs. In light of those results and the fact that PGAS-BPF with $N=1000$ takes substantially more CPU time than PGAS-PEIS with $N=30$ (224.7 versus 132.9 min) we report in Table 5 the posterior parameter estimates only for the PGAS-PEIS ($N=30$).  All parameter estimates are reasonable. The estimates of the parameters ($\delta_s$,$\sigma_s$) reveal that the state processes exhibit substantial variation and strong persistence, which is in full accordance  with the results reported by studies of realized (co)variances. The ESS values indicate that even for a challenging specification with as many as 26 parameters and 12,570 state variables the PGAS based on the PEIS  is capable to  produce numerically accurate and reliable parameter estimates with a fairly moderate computing time.

\section*{7. Conclusions}
The particle Gibbs (PG) is a flexible and easy to implement tool for conducting Bayesian analyses  of state space models. It uses sequential Monte Carlo (SMC) inside the Gibbs procedure in order to update the latent state trajectories. However, in high-dimensional applications when there is path degeneracy in the underlying SMC sampler the baseline PG suffers from severe  mixing problems.
Refinements designed to improve the mixing of the baseline PG introduce an ancestor sampling step to the underlying SMC (PGAS) or an additional Metropolis-Hastings move for the  update of the state trajectories (PGMH). However, such refinements when implemented using a standard locally designed SMCs such as the bootstrap particle filter of Gordon et al.~(1993) can still be prone to mixing problems, particularly, in applications involving narrowly distributed measurement variables coupled with outliers.

Here, we have proposed to combine the PG and its refinements with Particle Efficient Importance Sampling (PEIS) to overcome the mixing problem of the PG. The PEIS is an SMC  based on a recursive sequence of simple auxiliary regressions designed to construct  SMC importance sampling densities and resampling weights, which are nearly perfectly globally adapted to the targeted posterior density of the states.     We have shown that the PG  when combined with PEIS leads to significant improvements of the mixing w.r.t.~the state trajectories relative to PG procedures  based on standard locally designed SMCs. By such improvements of  the mixing, PG implementations based on  PEIS allow for numerically accurate and reliable Bayesian parameter  estimates not only in univariate state space models but also in multivariate high-dimensional specifications  as illustrated by the applications to a stochastic volatility model for asset returns, a square-root diffusion model for interest rates and an inverted Wishart model for the  realized covariance matrix of asset returns.
\section*{Acknowledgements}
We thank two anonymous referees for their helpful and constructive comments. We also thank participants of the 2015 Rhenisch Multivariate Time Series Econometrics (RMSE) Meeting (University of Cologne), the 2015 CMStatistics (ERCIM) conference, and the 2017 European Meeting of the Econometric Society (ESEM).  We are grateful to the Regional Computing Center at the University of Cologne for providing parts of the computational resources required.
\section*{References}
\begin{small}
\begin{description}
\item[]A\"{\i}t-Sahalia, Y., 1999. Transition densities for interest rate and other nonlinear diffusions. Journal of Finance 54, 1361-1395.
\item[]Anderson, T.W., 1984. An Introduction to Multivariate Statistical Analysis. John Wiley \& Sons, New York.
\item[]Andrieu, C., Doucet, A., and Holenstein, R., 2010. Particle Markov Chain Monte Carlo methods. Journal of the Royal Statistical Society 72, Series  B,  269-342.
\item[]Barndorff-Nielsen, O.E, Shephard, N., 2004. Econometric analysis of realized covariation: high frequency based covariance, regression, and correlation in financial economics. Econometrica 72, 885-925.
\item[]Black, F., 1995. Interest rates as options. Journal of Finance 50,  1371-1376.
\item[]Capp\'{e}, O., Godsill, S.J., and Moulines, E., 2007. An overview of exsting methods and recent advances in sequential Monte Carlo. Proceedings of the IEEE 95, 899-924.
\item[]Cox, J.C., Ingersoll, J.E., and Ross, S.A., 1985. A theory of the term structure of interest rates. Econometrica 53, 385-407.
 The Review of Economic Sudies 80, 538-567
\item[]Doucet, A., and Johansen, A.M., 2009. A tutorial on particle filtering and smoothing: Fifteen years later.
       In: Crisan, D., Rozovskii, B. (eds), The Oxford Handbook of Nonlinear Filtering. Oxford University Press, 656-704.
\item[]Flury, T., and Shephard, N., 2011. Bayesian inference based only on simulated likelihood: Particle filter analysis of dynamic economic models. Econometric Theory 27,  933-956.
\item[]Geyer, C.J., 1992. Practical Markov Chain Monte Carlo. Statistical Science 7,  473-483.
\item[]Ghysels, E., Harvey, A., and Renault, E., 1996. Stochastic Volatility. In:  Maddala, G., Rao, C.R. (eds), Handbook of Statistics, Vol 14.  Elsevier Sciences, 119-191.
\item[]Golosnoy, V., Gribisch, B., Liesenfeld, R., 2012. The conditional autoregressive Wishart model for multivariate stock market volatility. Journal of Econometrics 167, 211-223.
\item[]Gordon, N.J., Salmond, D.J., and Smith, A.F.M., 1993. A novel approach to non-linear and non-Gaussian Bayesian state estimation. IEEE Proceedings-F 140, 107-113.
\item[]Gorovoi, V., Linetsky, V., 2004. Black's model of interest rates as options, eigenfunction expansions an Japanese interest rates. Mathematical Finance 14, 49-78.
\item[]Guarniero, P., Johansen, A.M., Lee, A., 2017. The iterated particle filter. Journal of the American Statistical Association, in press (DOI:10.1080/01621459.2016.1222291).
\item[]Holenstein, R., 2009. Particle Markov Chain Monte Carlo. PhD-Thesis, University of British Colombia.
\item[]Kim, S., Shephard, N., and Chib, S., 1998. Stochastic volatility:
       Likelihood inference and comparison with ARCH models. Review of Economic Studies 65, 361-393.
\item[]Lin, M., Chen, R., and Liu, J.S., 2013. Lookahead strategies for sequential Monte Carlo. Statistical Science 28, 69-94.
\item[]Lindsten, F., Jordan, M.I., and Sch\"on, T.B., 2014. Particle Gibbs with ancestor sampling. Journal of Machine Learning Research 15, 2145-2184.
\item[]Lindsten, F., and Sch\"on, T.B., 2012. On the use of backward simulation in particle
      Markov chain Monte Carlo methods. Working paper, Link\"oping University, Sweden.
\item[]Liu, J.S., 1994. The collapsed Gibbs sampler in Bayesian computations with a applications to a gene regulation problem. Journal of the American Statistical Association 89,  958-966.
\item[]Pitt, M.K., Silva, d.S.R., Giordani, P., Kohn, R., 2012. On some properties of Markov chain Monte Carlo simulation methods based on the particle filter. Journal of Econometrics 171, 134-151.
\item[]Pitt, M.K., Shephard, N., 1999. Filtering via simulation: Auxiliary particle filters. Journal of the American Statistical Association 94, 590-599.
\item[]Richard, J.-F., Zhang, W., 2007. Efficient high-dimensional importance sampling. Journal of Econometrics 141, 1385-1411.
\item[]Scharth, M., and Kohn, R., 2016. Particle Efficient Importance Sampling. Journal of Econometrics 190, 133-147.
\item[]Tsay, R.S., 2010. Analysis of Financial Time Series. John Wiley \& Sons, Hoboken, New Jersey.
\item[]Whiteley, N., 2010.  Discussion on: Particle Markov Chain Monte Carlo methods. Journal of the Royal Statistical Society 72, Series  B,  306-307.
\item[]Whiteley, N., Andrieu, C., Doucet, A., 2010. Efficient Bayesian inference for switching state space models using discrete particle  Markov chain Monte Carlo methods. Bristol Statistics Research Report 10:04, University of Bristol.
\end{description}
\end{small}


\pagebreak

\begin{figure}[!htb]
\includegraphics[width=170mm,angle=0]{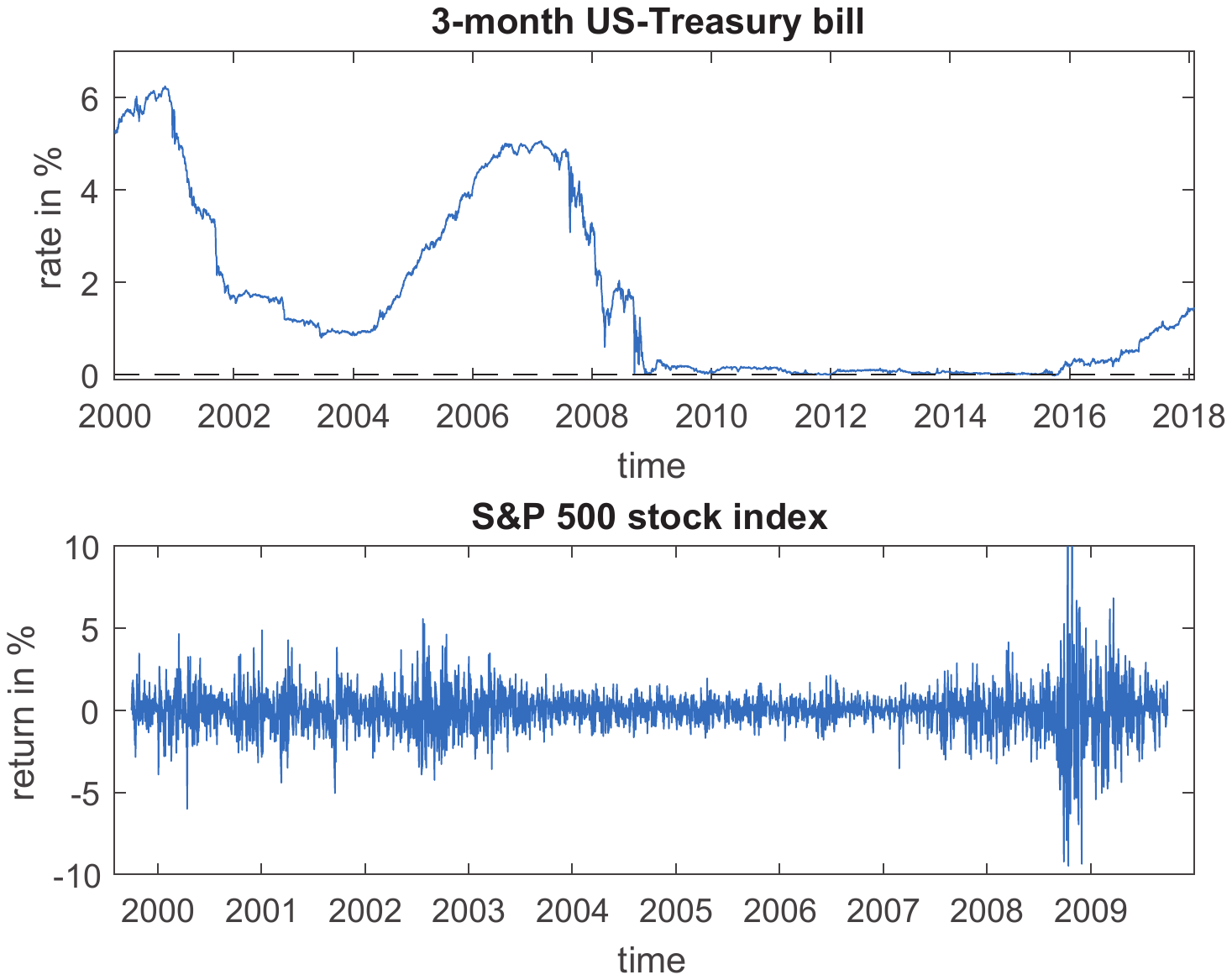}
\end{figure}
\vspace*{-0.0cm}
\begin{small}
\begin{center}
{\sl Figure 1. Top panel: The daily 3-month US-Treasury bill  rates from 2000 to 2018, the dashed line indicates the origin;
Bottom panel:  The daily returns on the S\&P 500 stock index from 1999 to 2009.}
\end{center}
\end{small}


\pagebreak

\begin{figure}[!htb]
\hspace{-1.2cm}\includegraphics[width=200mm,angle=0]{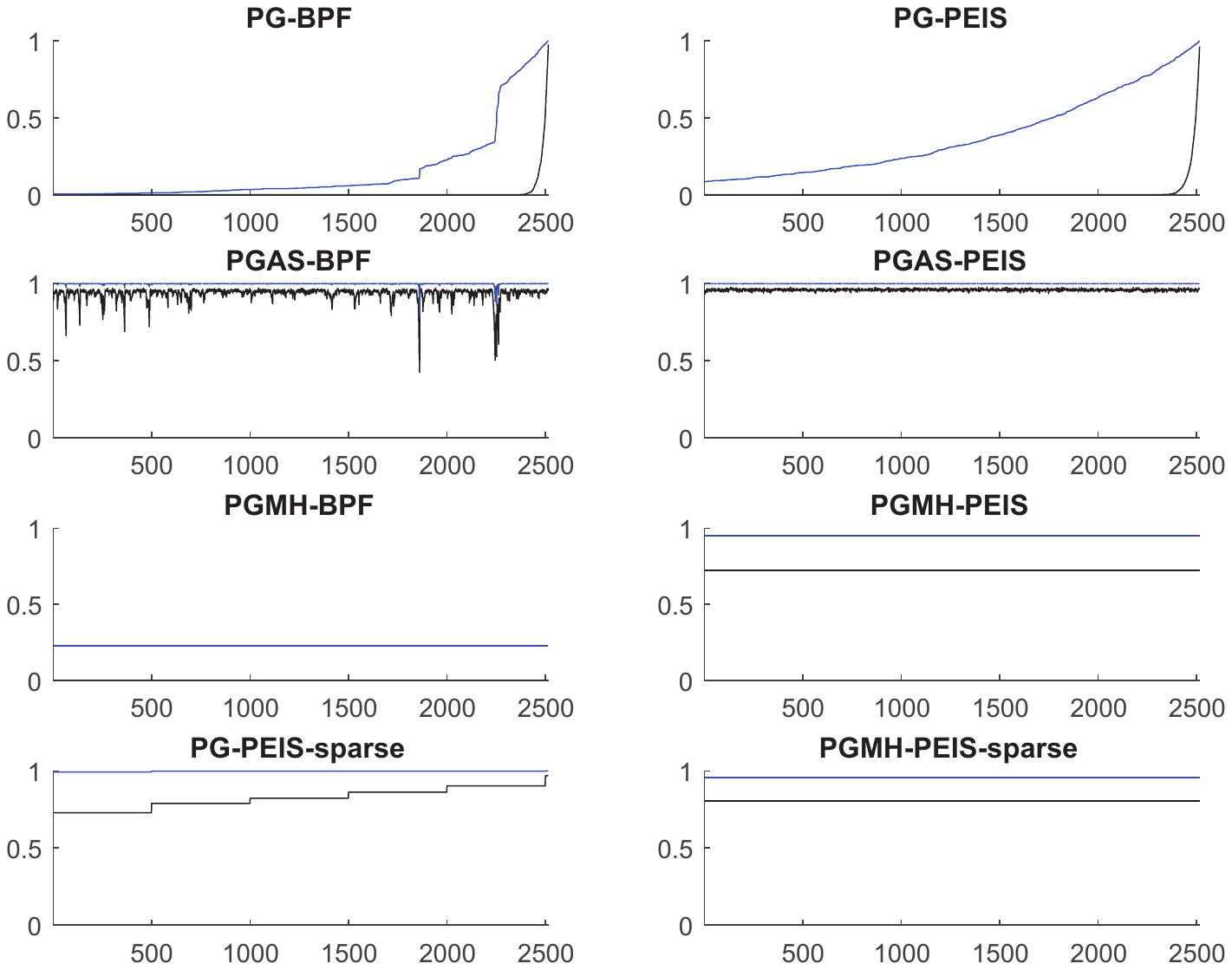}
\end{figure}
\vspace*{-0.0cm}
\begin{small}
\begin{center}
{\sl Figure 2. PG update rates for $x_t$  versus $t=1,...,T$  for the SV model, using $N=30$ particle (black line) and $N=1000$ (blue line). The ideal rates are $(N-1)/N$ for the PG and PGAS procedures and 1 for the PGMH procedures.}
\end{center}
\end{small}


\pagebreak

\begin{figure}[!htb]
\hspace{-1.2cm}\includegraphics[width=200mm,angle=0]{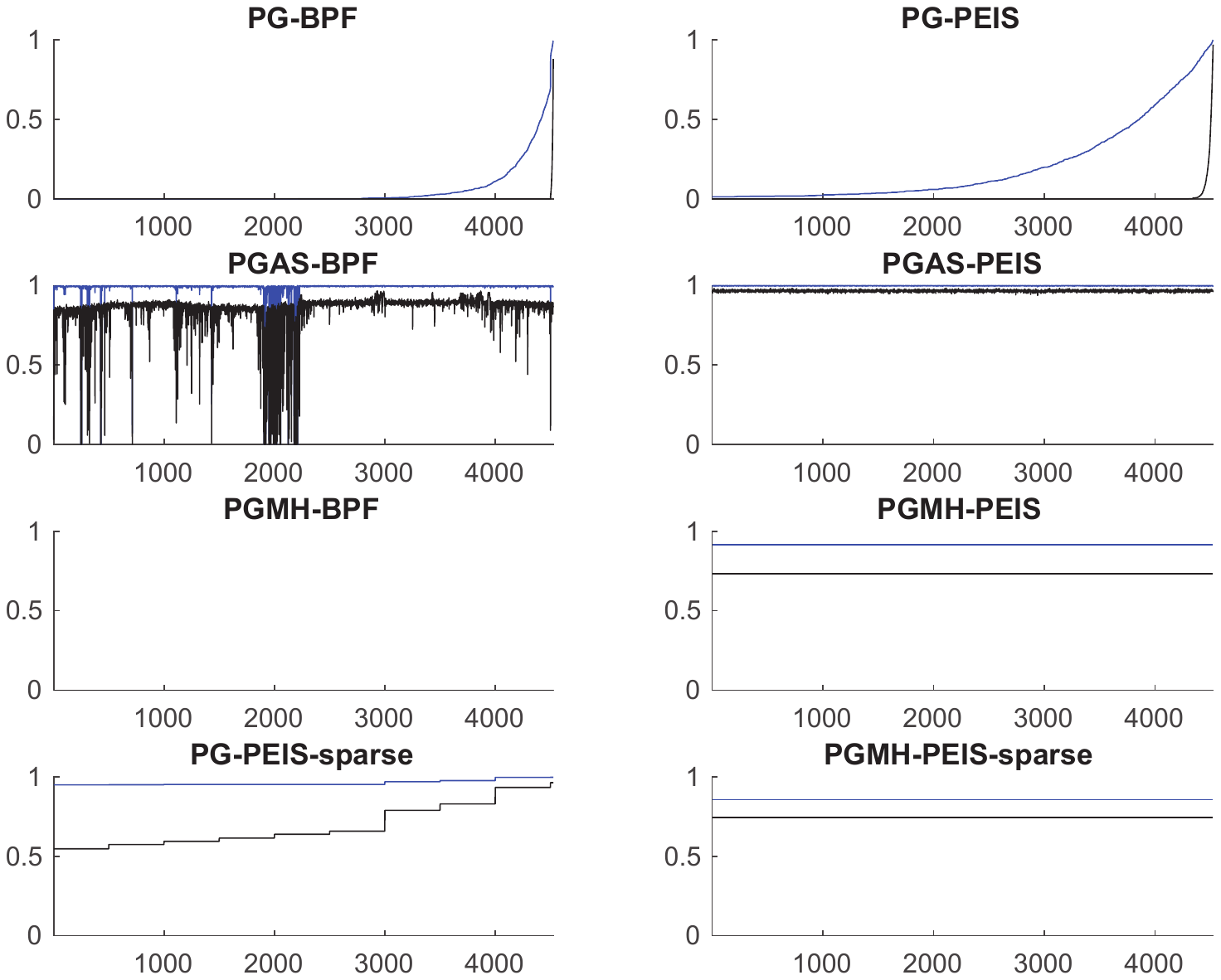}
\end{figure}
\begin{small}
\begin{center}
{\sl Figure 3. PG update rates for $x_t$  versus $t=1,...,T$  for the shifted CIR model, using $N=30$ particle (black line) and $N=1000$ (blue line). The ideal rates are $(N-1)/N$ for the PG and PGAS procedures and 1 for the PGMH procedures.}
\end{center}
\end{small}


\pagebreak

\vspace*{-2.0cm}
\begin{center}
\begin{figure}[!htb]
\hspace*{1.5cm}\includegraphics[width=130mm,angle=0]{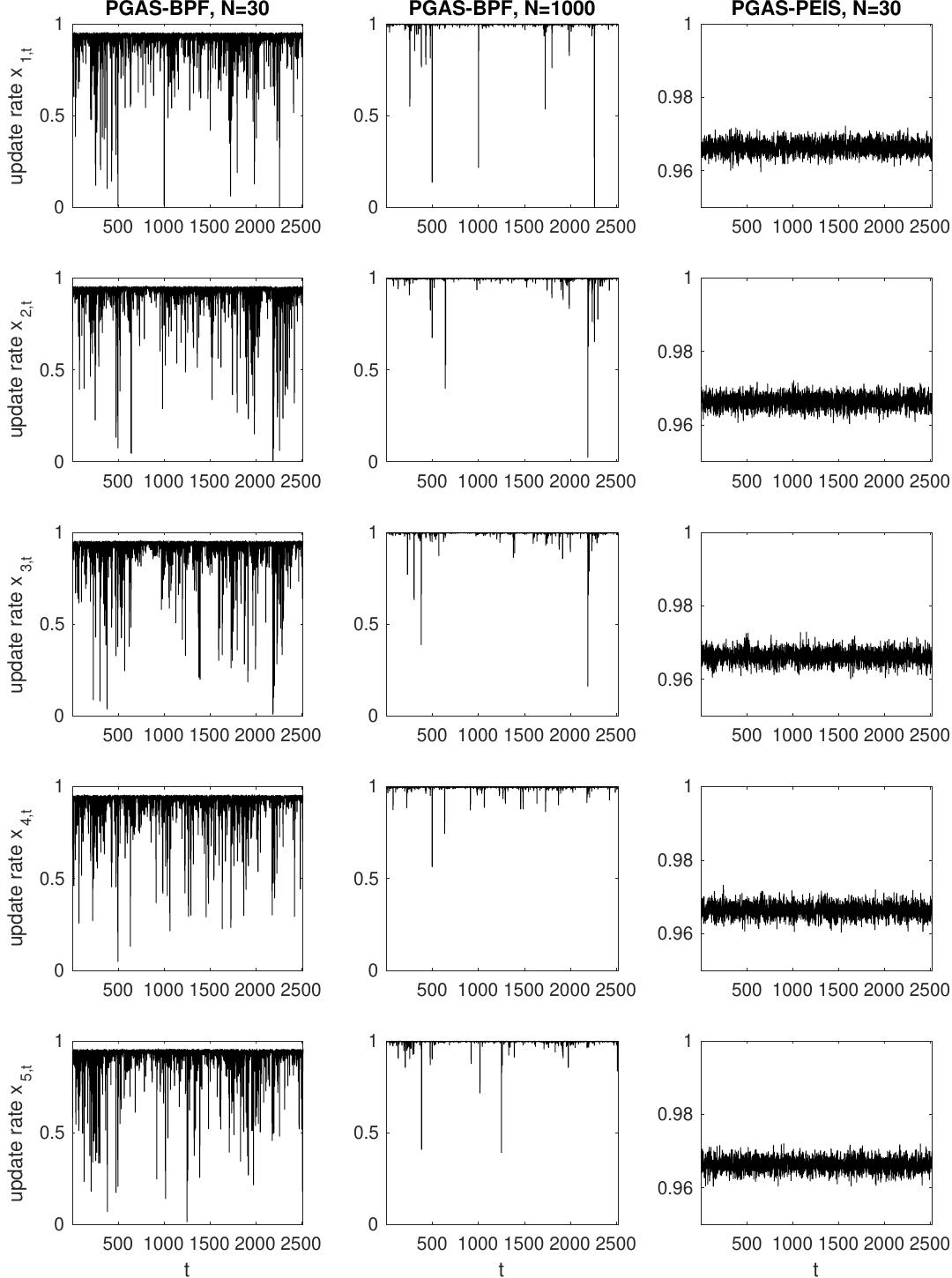}
\end{figure}
\end{center}
\vspace*{-1.0cm}
\begin{small}
\begin{center}
{\sl Figure 4. PGAS update rates for $x_t = (x_{1t},\ldots,x_{5t})$  versus $t=1,...,T$  for the inverted Wishart model under the BPF with $N=30$ (left panel), BPF with $N=1000$ (middle panel), PEIS with $N=30$ (right panel). The ideal rates are $(N-1)/N$.}
\end{center}
\end{small}

\onecolumn
\newcolumntype{.}{D{.}{.}{3}}


\begin{table}\centering
\footnotesize{
\begin{tabular}{lrrrrrr}
\multicolumn{7}{c}{\mbox{\small{\sl Table 1. Effective Sample Size for PG Samples from the Posterior of the States }}}\\
\multicolumn{7}{c}{\mbox{\small{\sl in the SV Model for Fixed Parameters}}}
\\[0.1cm]\hline\\[-0.3cm]
& \multicolumn{1}{c}{Number of} & \multicolumn{1}{c}{CPU time}  & \multicolumn{1}{c}{Minimum} &\multicolumn{1}{c}{Median} &\multicolumn{1}{c}{Maximum}&\multicolumn{1}{c}{Minimum}\\
& \multicolumn{1}{c}{particles} & \multicolumn{1}{c}{in sec}    & \multicolumn{1}{c}{ESS}     &\multicolumn{1}{c}{ESS}    &\multicolumn{1}{c}{ESS}&\multicolumn{1}{c}{ESS per }\\
&                               &                               &                             &                           &                       &\multicolumn{1}{c}{hour CPU }\\
&                               &                               &                             &                           &                       &\multicolumn{1}{c}{time}\\
\hline\\
PG-BPF            &  30   &   283  &   1   & 1  & 621   &  13  \\
PG-BPF            &  1000 &   942  &   3   & 30 & 1000  &  12  \\

PG-PEIS           &  30   &   1646 &   1   & 1  & 566   &   2  \\
PG-PEIS           &  1000 &   2397 &   21  & 173& 999   &   31  \\

PG-PEIS-sparse    &  30   &   1552 &   332 & 671& 969   &  771 \\
PG-PEIS-sparse    &  1000 &   2249 &   569 & 948& 1000  &  912 \\[0.2cm]

PGAS-BPF          &  30   &   653  &   45  & 415& 689   &  254  \\
PGAS-BPF          &  1000 &   1499 &   297 & 934& 1000  &  716  \\
PGAS-PEIS         &  30   &   2042 &   240 & 475& 707   &  423\\
PGAS-PEIS         &  1000 &   3038 &   573 & 949& 1000  &  686\\[0.2cm]

PGMH-BPF          &  30   &   881  &   1   & 1  &  1    &  4  \\
PGMH-BPF          &  1000 &   2454 &   34  & 113& 210   &  51  \\
PGMH-PEIS         &  30   &   2313 &   284 & 538& 756   &  442 \\
PGMH-PEIS         &  1000 &   3798 &   532 & 876& 1000  &  505 \\
PGMH-PEIS-sparse  &  30   &   1906 &   355 & 662& 860   &  674 \\
PGMH-PEIS-sparse  &  1000 &   3345 &   552 & 894& 1000  &  563 \\[0.2cm]
\hline
\end{tabular}
}
\begin{footnotesize}
\begin{quote}
{\sl NOTE: Results for the PG algorithms are based on 1,100 PG iterations (discarding the first 100 draws). All reported statistics are sample averages computed from 10 independent replications of the PG algorithms under 10 different seeds.}
\end{quote}
\end{footnotesize}
\end{table}





\begin{table}\centering
\footnotesize{
\begin{tabular}{lrrrrrr}
\multicolumn{7}{c}{\mbox{\small{\sl Table 2. Effective Sample Size for PG Samples from the Posterior of the States }}}\\
\multicolumn{7}{c}{\mbox{\small{\sl in the shifted CIR model for Fixed Parameters}}}
\\[0.1cm]\hline\\[-0.3cm]
& \multicolumn{1}{c}{Number of} & \multicolumn{1}{c}{CPU time}  & \multicolumn{1}{c}{Minimum} &\multicolumn{1}{c}{Median} &\multicolumn{1}{c}{Maximum}&\multicolumn{1}{c}{Minimum}\\
& \multicolumn{1}{c}{particles} & \multicolumn{1}{c}{in sec}    & \multicolumn{1}{c}{ESS}     &\multicolumn{1}{c}{ESS}    &\multicolumn{1}{c}{ESS}&\multicolumn{1}{c}{ESS per }\\
&                               &                               &                             &                           &                       &\multicolumn{1}{c}{hour CPU }\\
&                               &                               &                             &                           &                       &\multicolumn{1}{c}{time}\\
\hline\\
PG-BPF            &  30   &   78   &   1   & 1  & 875   &  47  \\
PG-BPF            &  1000 &   481  &   1   & 2  & 996   &   8  \\

PG-PEIS           &  30   &   2342 &   1   & 1  & 953   &   2  \\
PG-PEIS           &  1000 &   4406 &   4   & 59 & 1000  &   3  \\

PG-PEIS-sparse    &  30   &   2292 &   93  & 407& 1000  &  145 \\
PG-PEIS-sparse    &  1000 &   3984 &   281 & 794& 1000  &   254 \\[0.2cm]

PGAS-BPF          &  30   &   140  &   1   & 840& 1000  &  26  \\
PGAS-BPF          &  1000 &   652  &   1   & 963& 1000  &   6  \\
PGAS-PEIS         &  30   &   2341 &   242 & 904& 1000  &  373\\
PGAS-PEIS         &  1000 &   4299 &   433 & 966& 1000  &   363\\[0.2cm]

PGMH-BPF          &  30   &   203  &   1   & 1  & 1     &   18  \\
PGMH-BPF          &  1000 &   1104 &   1   & 1  & 1     &   3  \\
PGMH-PEIS         &  30   &   2605 &   165 & 529& 793   &  227 \\
PGMH-PEIS         &  1000 &   6204 &   319 & 808& 1000  &   186 \\
PGMH-PEIS-sparse  &  30   &   2605 &    85 & 415&  753  &  117 \\
PGMH-PEIS-sparse  &  1000 &   6065 &   246 & 737&  995  &   146 \\[0.2cm]
\hline
\end{tabular}
}
\begin{footnotesize}
\begin{quote}
{\sl NOTE: Results for the PG algorithms are based on 1,100 PG iterations (discarding the first 100 draws). All reported statistics are sample averages computed from 10 independent replications of the PG algorithms under 10 different seeds.}
\end{quote}
\end{footnotesize}
\end{table}

\vspace*{1cm}

\newpage


\begin{table}\centering
\footnotesize{
\begin{tabular}{llrrrrrr}
\multicolumn{8}{c}{\mbox{\small{\sl Table 3. PG Posterior Analysis of the SV Model }}}\\
\\[0.1cm]\hline\\[-0.3cm]
&& \multicolumn{1}{c}{PG-}         & \multicolumn{1}{c}{PGAS-}  & \multicolumn{1}{c}{PGAS-}    &\multicolumn{1}{c}{PGMH-}   &\multicolumn{1}{c}{PGMH-}&\multicolumn{1}{c}{Ideal}       \\
&& \multicolumn{1}{c}{PEIS-}       & \multicolumn{1}{c}{BPF}    & \multicolumn{1}{c}{PEIS}     &\multicolumn{1}{c}{PEIS}    &\multicolumn{1}{c}{PEIS-}&\multicolumn{1}{c}{Gibbs}\\
&& \multicolumn{1}{c}{sparse}      & \multicolumn{1}{c}{}       & \multicolumn{1}{c}{}         &\multicolumn{1}{c}{}        &\multicolumn{1}{c}{sparse}&\\
\hline\\
               &   CPU time (hours)    &      14:30   &  5:49    &  18:45     & 21:26   & 17:15     &   \\
\hline\\
$\beta\qquad$  &   post. mean          &     1.0617   &  1.0645  &   1.0754   & 1.0580  & 1.0727   & 1.0708 \\
               &   post. std.          &     0.1855   &  0.1943  &   0.1892   & 0.2214  & 0.1978   & 0.2003  \\
               &   ESS               &     96       &  41      &    77      & 72      & 96       & 112\\
               & ESS/hour CPU time   &     7        &  7       &     4      & 3       & 6        &  \\[0.2cm]
$\delta$       &   post. mean          &     0.9924   &  0.9924  &   0.9924   & 0.9926  & 0.9924   & 0.9924 \\
               &   post. std.          &     0.0027   &  0.0027  &   0.0027   & 0.0028  & 0.0027   & 0.0028 \\
               &   ESS               &     653      &  467     &   582      & 474     & 538      & 694\\
               & ESS/hour CPU time   &     45       &  80      &   31       & 22      & 31       & \\[0.2cm]
$\nu$          &   post. mean          &     0.1205   &  0.1204  &   0.1201   & 0.1203  & 0.1202   & 0.1206 \\
               &   post. std.          &     0.0125   &  0.0125  &   0.0125   & 0.0123  & 0.0125   & 0.0128 \\
               &   ESS               &     265      &  345     &    355     & 226     & 251      & 356  \\
               & ESS/hour CPU time   &     18       &  59      &    19      &  11     & 15      & \\[0.2cm]
\hline
\end{tabular}
}
\begin{footnotesize}
\begin{quote}
{\sl NOTE: Results for the PG algorithms are based on 50,000 PG iterations (discarding the first 10,000 draws) and $N=30$ SMC particles. All reported statistics are sample averages computed from 10 independent replications of the PG algorithms under 10 different seeds. The ML estimates for the parameters are $(\beta,\delta,\nu)$ = (1.065, 0.992, 0.122).}
\end{quote}
\end{footnotesize}
\end{table}


\begin{table}\centering
\footnotesize{
\begin{tabular}{llrrrrrr}
\multicolumn{8}{c}{\mbox{\small{\sl Table 4. PG Posterior Analysis of the shifted CIR Model }}}\\
\\[0.1cm]\hline\\[-0.3cm]
&& \multicolumn{1}{c}{PG-}         & \multicolumn{1}{c}{PGAS-}  & \multicolumn{1}{c}{PGAS-}    &\multicolumn{1}{c}{PGMH-}   &\multicolumn{1}{c}{PGMH-}&\multicolumn{1}{c}{Ideal}\\
&& \multicolumn{1}{c}{PEIS-}       & \multicolumn{1}{c}{BPF}    & \multicolumn{1}{c}{PEIS}     &\multicolumn{1}{c}{PEIS}    &\multicolumn{1}{c}{PEIS-}&\multicolumn{1}{c}{Gibbs}\\
&& \multicolumn{1}{c}{sparse}      & \multicolumn{1}{c}{}       & \multicolumn{1}{c}{}         &\multicolumn{1}{c}{}        &\multicolumn{1}{c}{sparse}&\\
\hline\\
               &   CPU time (hours)    &      31:17   &  1:51   &     33:42   & 38:13   & 36:15      &\\
\hline\\
$\alpha\qquad$ &   post. mean          &     0.0013   &  0.0014  &   0.0013   & 0.0013  & 0.0013     & 0.0013\\
               &   post. std.          &     0.0021   &  0.0021  &   0.0021   & 0.0021  & 0.0021     & 0.0021\\
               &   ESS                 &     39795    &  --      &    39532   & 39621   & 39624      & 39345\\
               & ESS/hour CPU time     &     1274     &  --      &    1173    & 1045    & 1094       &\\[0.2cm]
$\beta$        &   post. mean          &     0.2180   &  0.2206  &   0.2180   & 0.2181  & 0.2183     & 0.2180\\
               &   post. std.          &     0.1048   &  0.1065  &   0.1045   & 0.1048  & 0.1047     & 0.1048\\
               &   ESS                 &     39584    &  --      &    39616   & 39535   & 39573      & 39705\\
               & ESS/hour CPU time     &     1267     &  --      &     1175   & 1043    & 1092       &\\[0.2cm]
$\sigma_x$     &   post. mean          &     0.0287   &  0.0292  &   0.0287   & 0.0287  & 0.0287     & 0.0287\\
               &   post. std.          &     0.0003   &  0.0003  &   0.0003   & 0.0003  & 0.0003     & 0.0003\\
               &   ESS                 &     14567    &  --      &    22596   & 18944   & 13557      & 24270 \\
               & ESS/hour CPU time     &     467      &  --      &     671    &  501    & 374        &\\[0.2cm]
$\sigma_y$     &   post. mean          &     9.8e-5   &  6.0e-5  &   9.8e-5   & 9.8e-5  & 9.8e-5     & 9.8e-5 \\
               &   post. std.          &     6.6e-6   &  3.7e-6  &   6.6e-6   & 6.6e-6  & 6.5e-6     &  6.5e-6\\
               &   ESS                 &     318      &  --      &    470     & 327     & 326        &  481\\
               & ESS/hour CPU time     &     10       &  --      &     14     &  9      & 9          & \\[0.2cm]
\hline
\end{tabular}
}
\begin{footnotesize}
\begin{quote}
{\sl NOTE: Results for the PG algorithms are based on 50,000 PG iterations (discarding the first 10,000 draws) and $N=30$ SMC particles. All reported statistics are sample averages computed from 10 independent replications of the PG algorithms under 10 different seeds.  The ML estimates for the parameters are $(\alpha,\beta,\sigma_x,\sigma_y)$ = (0.0013, 0.2179, 0.0287, 9.8e-5).}
\end{quote}
\end{footnotesize}
\end{table}


\begin{table}\centering
\small{
\begin{tabular}{lrrrrrrrrrr}
\multicolumn{11}{c}{\mbox{\small{\sl Table 5. PGAS-PEIS Posterior Analysis of the Inverted-Wishart Model}}}\\
\\[0.1cm]\hline\\[-0.3cm]
               &   $\mu_1$ &   $\mu_2$ &   $\mu_3$ &   $\mu_4$ &   $\mu_5$ & $\delta_1$ & $\delta_2$& $\delta_3$& $\delta_4$& $\delta_5$   \\
\hline\\
post. mean     &      4.15 &   4.12    &  3.72     & 4.11      &    3.53   & 0.97       &  0.98     & 0.96      &  0.94     &  0.96        \\
post. std.     &      0.20 &   0.26    &  0.15     & 0.10      &    0.13   & 0.005      &  0.004    & 0.006     &  0.008    &  0.006        \\
ESS            &      9815 &   9731    &  9369     & 9326      &    9571   & 4101       &  4622     & 4162      &  2904     &  3159       \\
\hline\\
               &   $\sigma_1$ &   $\sigma_2$ &   $\sigma_3$ &   $\sigma_4$ &   $\sigma_5$ & $\nu$   &  & & &    \\
\hline\\
post. mean     &      0.31 &   0.26    &  0.29     & 0.28      &    0.25   & 33.6                   &  & & &    \\
post. std.     &      0.009&   0.008   &  0.009    & 0.009     &    0.009  & 0.28                   &  & & &    \\
ESS            &      1044 &   1032    &  1053     & 944       &    809    & 275                     &  & & &    \\
\hline\\
               &   $\tilde h_{1,1}$ &  $\tilde h_{1,2}$ & $\tilde h_{1,3}$  & $\tilde h_{1,4}$ & $\tilde h_{2,1}$ & $\tilde h_{2,2}$ & $\tilde h_{2,3}$ & $\tilde h_{3,1}$ & $\tilde h_{3,2}$ & $\tilde h_{4,1}$ \\
\hline\\
post. mean     &      0.39 &   0.29    &  0.29     & 0.23      &    0.20   & 0.17       &  0.12     & 0.22      &  0.18     &  0.11        \\
post. std.     &      0.003&   0.003   &  0.003    & 0.002     &    0.003  & 0.003      &  0.002    & 0.004     &  0.003    &  0.002        \\
ESS            &      8260 &   8566    &  8860     & 8398      &    8530   & 7828       &  7274     & 9173      &  9046     &  9413       \\
\hline
\end{tabular}
}
\begin{footnotesize}
\begin{quote}
{\sl NOTE: Results are based on 15,000 PGAS-PEIS iterations (discarding the first 5,000 draws) and $N=30$ SMC particles. All reported statistics are sample averages computed from 10 independent replications of the PGAS-PEIS algorithms under 10 different seeds.}
\end{quote}
\end{footnotesize}
\end{table}

\end{document}